\documentclass[%
     aip,
     amsmath,amssymb,
     reprint,%
]{revtex4-1}

\usepackage{graphicx}% Include figure files
\usepackage{dcolumn}% Align table columns on decimal point
\usepackage{bm}% bold math
\usepackage[utf8]{inputenc}
\usepackage[T1]{fontenc}
\usepackage{mathptmx}
\usepackage{etoolbox}

\makeatletter
\def\@email#1#2{%
 \endgroup
 \patchcmd{\titleblock@produce}
  {\frontmatter@RRAPformat}
  {\frontmatter@RRAPformat{\produce@RRAP{*#1\href{mailto:#2}{#2}}}\frontmatter@RRAPformat}
  {}{}
}%
\makeatother

\usepackage[margin=2cm]{geometry}
\allowdisplaybreaks

\usepackage{subfig}

\usepackage{hyperref}
\usepackage[dvipsnames]{xcolor}
\usepackage{amsmath}
\usepackage{amsfonts,amssymb}

\usepackage{tikz}
\usetikzlibrary{quantikz2}
\usepackage{arydshln}
\usepackage{soul}  % to use the strikeout text

\usepackage{pgfplots}
\pgfplotsset{compat=1.18}

\usepackage{mathbbol}
\usepackage{xifthen}

\usepackage[titletoc,title]{appendix}

\usepackage{hhline}
\usepackage{multirow}

\colorlet{ygray}{gray!20}

\newcommand{\aeq}{\begin{equation}}
\newcommand{\eeq}{\end{equation}}
\newcommand{\aeqn}{\begin{eqnarray}}
\newcommand{\eeqn}{\end{eqnarray}}
\newcommand{\aeqns}{\begin{eqnarray*}}
\newcommand{\eeqns}{\end{eqnarray*}}

\newcommand{\yi}{\mathrm{i}}

\newcommand{\oO}{\mathcal{O}}
\newcommand*\diff{\mathop{}\!\mathrm{d}}
\newcommand{\itimes}[2]{\langle #1|#2 \rangle}

\newcommand{\ylb}{\left(}
\newcommand{\yrb}{\right)}

\newcommand{\yEx}[1][]{\ifthenelse{\isempty{#1}}{\tilde{E}_x}{\tilde{E}_{x,#1}}}
\newcommand{\yEy}[1][]{\ifthenelse{\isempty{#1}}{\tilde{E}_y}{\tilde{E}_{y,#1}}}
\newcommand{\yBz}[1][]{\ifthenelse{\isempty{#1}}{\tilde{B}_z}{\tilde{B}_{z,#1}}}

\newcommand{\yff}{\textcolor{black}{f}}
\newcommand{\ypf}{\textcolor{black}{g}}
\newcommand{\yfE}{\textcolor{black}{E}}
\newcommand{\ypE}{\textcolor{black}{E}}
\newcommand{\yBF}{\textcolor{black}{F}}

\newcommand{\lambdaD}{\lambda_{\rm D}}
\newcommand{\omegaP}{\omega_{\rm p}}
\newcommand{\yvth}{v_{\rm th}}
\newcommand{\ynref}{n_{\rm ref}}
\newcommand{\yTref}{T_{\rm ref}}
\newcommand{\yme}{m_e}

\newcommand{\yjs}{j^{(S)}}
\newcommand{\yrhos}{\rho^{(S)}}

\newcommand{\yLf}{\ypf_\omega}
\newcommand{\yLE}{\ypE_\omega}
\newcommand{\yLj}{\yjs_\omega}

\newcommand{\yLS}{S_\omega}
\newcommand{\yLFf}{\ypf_{\omega, k}}
\newcommand{\yLFE}{\ypE_{\omega,k}}
\newcommand{\yLFj}{\yjs_{\omega,k}}
\newcommand{\yLFrho}{\yrhos_{\omega,k}}
\newcommand{\yFEi}{\ypE_{0,k}}
\newcommand{\yLFS}{S_{\omega,k}}
\newcommand{\yFQ}{Q_{k}}

\newcommand{\yFE}{\ypE_{k}}
\newcommand{\yFrho}{\yrhos_{0,k}}

\newcommand{\ymatrixstyle}{\mathbf}
\newcommand{\yP}{\ymatrixstyle{P}}
\newcommand{\yU}{\ymatrixstyle{U}}
\newcommand{\yA}{\ymatrixstyle{A}}

\newcommand{\yF}{\ymatrixstyle{F}}
\newcommand{\yC}{\ymatrixstyle{C}}
\newcommand{\yS}{\ymatrixstyle{S}}
\newcommand{\yD}{\ymatrixstyle{D}}

\newcommand{\yG}{\ymatrixstyle{G}}
\newcommand{\yM}{\ymatrixstyle{M}}

\newcommand{\yFF}{\hat{\ymatrixstyle{F}}}
\newcommand{\ysFF}{\hat{F}}

\newcommand{\yFP}{\tilde{\ymatrixstyle{F}}}

\newcommand{\ypsi}{\pmb{\psi}}
\newcommand{\yyb}{\boldsymbol{b}}

\newcommand{\ydiag}{\mathcal{D}}
\newcommand{\yset}{S}
\newcommand{\ysete}{\yset^{\rm ext}}
\newcommand{\ysetc}{\yset^{\rm com}}

\newcommand{\yvset}{v^{(S)}}

\newcommand{\yoset}[1]{\overrightarrow{\{r\}}_{#1}}

\newcommand{\ycite}{\cite}
\newcommand{\yocite}{\onlinecite}

\usepgfplotslibrary{fillbetween}

\begin{document}
\preprint{AIP/123-QED}

%% -------------------------------------------------------
%% --- TITLE
%% -------------------------------------------------------
% If necessary make line breaks with \\
% \title[]{Encoding a kinetic model of anomalous skin effect into a quantum circuit}
% \title[]{Matrix compression for encoding kinetic plasma problems into quantum circuits}
\title[]{Encoding of linear kinetic plasma problems in quantum circuits via data compression}
\author{I. Novikau}
\email{novikau1@llnl.gov}
\affiliation{Lawrence Livermore National Laboratory, Livermore, California 94550, USA}
\author{I. Y. Dodin}
\affiliation{Princeton Plasma Physics Laboratory, Princeton, New Jersey 08543, USA}
\affiliation{Department of Astrophysical Sciences, Princeton University, Princeton, New Jersey 08544, USA}
% \author{I. Joseph}
% \affiliation{Lawrence Livermore National Laboratory, Livermore, California 94550, USA}
\author{E. A. Startsev}
\affiliation{Princeton Plasma Physics Laboratory, Princeton, New Jersey 08543, USA}

\date{\today}

%% -------------------------------------------------------
%% --- Abstract ---
%% -------------------------------------------------------
\begin{abstract}
We propose an algorithm for encoding of linear kinetic plasma problems in quantum circuits. 
The focus is made on modeling of electrostatic linear waves in one-dimensional Maxwellian electron plasma. 
The waves are described by the linearized Vlasov–Amp\`ere system with a spatially localized external current that drives plasma oscillations. 
This system is formulated as a boundary-value problem and cast in the form of a linear vector equation $\yA\ypsi = \yyb$ to be solved by using the quantum signal processing algorithm. 
The latter requires encoding of the matrix $\yA$ in a quantum circuit as a subblock of a unitary matrix.  
We propose how to encode $\yA$ in a circuit in a compressed form and discuss how the resulting circuit scales with the problem size and the desired precision.
\end{abstract}

%% -------------------------------------------------------
%% --- Bulk text ---
%% -------------------------------------------------------
\maketitle

\section{Introduction}\label{sec:introduction}

% ------------------------------------------------------------------------------------
Modeling of the plasma dynamics typically involves operating with classical fields on fine grids. 
This requires dealing with large amounts of data, especially in kinetic models, which are notorious for being computationally expensive. 
Quantum computing (QC) has the potential to significantly speed up kinetic simulations by leveraging quantum superposition and entanglement\ycite{Nielsen10}. 
However, quantum speedup is possible only if the depth of a quantum circuit modeling the plasma dynamics scales advantageously (polylogarithmically) with the system size (number of grid cells). 
Achieving such efficient encoding is challenging and remains an open problem for most plasma systems of practical interest. 

Here, we explore the possibility of an efficient quantum algorithm for modeling of linear oscillations and waves in a Vlasov plasma\ycite{Stix92}. 
Previous works in this area focused on modeling either spatially monochromatic or conservative waves within initial-value problems\ycite{Engel19, Ameri23, Toyoizumi23}. 
However, typical practical applications (for example, for magnetic confinement fusion) require modeling of inhomogeneous dissipative waves within boundary-value problems, which require different approaches. 
Here, we consider a minimal problem of this kind to develop a prototype algorithm that would be potentially extendable to such practical applications.

Specifically, we assume one-dimensionless collisionless Maxwellian electron plasma governed by the linearized Vlasov--Amp\`ere system. 
Added to this system is a spatially localized external current (antenna) that drives plasma oscillations at a fixed frequency $\omega$. 
This source produces evanescent waves if $\omega$ is smaller than the plasma frequency $\omega_p$. At $\omega > \omega_p$, it produces Langmuir waves, which propagate away from the source while experience Landau damping along the way. 
Outgoing boundary conditions are adopted to introduce irreversible dissipation. 
This relatively simple system captures paradigmatic dynamics typical for linear kinetic plasma problems and thus can serve as a testbed for developing more general algorithms of practical interest.
We start by showing how said system can be cast in the form of a linear vector equation\footnote{In this paper, matrices and vectors are indicated by bold symbols, as in Eq.~\eqref{eq:axb}, while quantum-state vectors, oracles, and operators in quantum circuits are denoted with non-bold italic symbols, e.g. $U$ and $\psi$.}
\begin{equation}\label{eq:axb}
    \yA \ypsi = \yyb,
\end{equation}
where $\yA$ is a (non-Hermitian) square matrix, $\ypsi$ is a vector that represents the dynamical variables (the electric field and the distribution function) on a grid, and $\yyb$ describes the antenna. 
There is a variety of quantum algorithms that can solve equations like~\eqref{eq:axb}, for example, the Harrow--Hassidim--Lloyd (HHL) algorithms\ycite{HHL09}, the Ambainis algorithm\ycite{Ambainis12}, the algorithms inspired by the adiabatic quantum computing\ycite{Costa21, Jennings23}, and solvers based on the Quantum Signal Processing (QSP)\ycite{Low17, Low19, Gilyen19, Martyn21}, to name a few.
Here, we propose to use a method based on the QSP, specifically, the Quantum Singular Value Transformation (QSVT)\ycite{Gilyen19, Martyn21}, because it is known to scale near-optimally with the condition number of $\yA$ and the desired precision.

Specifically, this paper focuses on the problem that one unavoidably has to overcome when applying the QSVT to kinetic plasma simulations. 
This problem is how to encode the corresponding large-dimensional matrix $\yA$ into a quantum circuit. 
A direct encoding of this matrix is prohibitively inefficient. 
Various methods for encoding matrices into quantum circuits were developed recently\ycite{Clader22, Camps23, Kuklinski24, Lapworth24, Liu24, Sunderhauf24, Zhang23}.
We propose how to make it more efficiently by compressing the content of $\yA$ at encoding. 
% We focus on encoding of the matrix $\yA$ into a quantum circuit, which the central part of any QSP-based algorithm.
% Since the matrix $\yA$ is not unitary, it cannot be directly decomposed into quantum gates.
% Instead, $\yA$ is encoded as a submatrix of an auxiliary unitary matrix $\yU_A$. 
% This procedure is known as block encoding (BE). 
% We explicitly show how to block encode the matrix $\yA$ specific to our electrostatic kinetic problem. 
The same technique can be applied in modeling kinetic or fluid plasma and electromagnetic waves in higher-dimensional problems, so our results can be used as a stepping stone towards developing more practical algorithms in the future. 
% We also discuss how our algorithm scales with the size of the problem and show that the encoding of $yA$ can be done efficiently. 
The presentation of the rest of the algorithm and (emulation of) quantum simulations are left to the future work.

Our paper is organized as follows. 
In Sec.~\ref{sec:theory}, we present the main equations of the electrostatic kinetic plasma problem. 
In Sec.~\ref{sec:discretization}, we present discretization of this model. 
In Sec.~\ref{sec:linear-equations}, we cast the discretized equations into the form~\eqref{eq:axb}. 
In Sec.~\ref{sec:classical-results}, we present classical numerical simulations of the resulting system, which can be used for benchmarking of the future quantum algorithm. 
In Sec.~\ref{sec:encoding}, we present the general strategy for the encoding of $\yA$ into a quantum circuit. 
In Sec.~\ref{sec:OH}, we explicitly construct the corresponding oracle and discuss how it scales with the parameters of the problem. 
In Sec.~\ref{sec:discussion}, we present a schematic circuit of the quantum algorithm based on this oracle. 
In Sec.~\ref{sec:conclusions}, we present our main conclusions.

\section{Model}\label{sec:theory}

% ----------------------------------------------------------------------------------------------
% ----------------------------------------------------------------------------------------------
\subsection{One-dimensional Vlasov--Amp\`ere system}\label{sec:theory-1D}
Electrostatic oscillations of one-dimensional electron plasma can be described by the Vlasov--Amp\`ere system
\begin{subequations}\label{sys:VM-init}
\begin{eqnarray}
    &&\partial_t \yff + v\partial_x \yff - \frac{e}{\yme}\yfE\partial_{v}\yff = F_{\rm ext},\\
    &&\partial_t \yfE - 4\pi e \int v \yff \diff v = - 4\pi \yjs,
\end{eqnarray}
\end{subequations}
where $\yfE(t, x)$ is the electric field, $\yff(t, x, v)$ is the electron probability distribution, $t$ is time, $x$ is the coordinate in the physical space, and $v$ is the coordinate in the velocity space.
$F_{\rm ext}$ is an external force of order $E^2$, e.g. the ponderomotive force due to an external laser radiation of a high frequency.
Also, $e > 0$ is the elementary change and $m_e$ is the electron mass. 
The term $\yjs$ represents a prescribed source current that drives plasma oscillations. 
The corresponding source charge density $\yrhos$ can be inferred from $\yjs$ using the charge conservation law:
\begin{equation}\label{eq:charge-conserv}
    \partial_t \yrhos + \partial_x\yjs = 0.
\end{equation}

Let us split the electron distribution into the background distribution $\yBF$ and a perturbation $\ypf$:
\begin{equation}
    \yff(t,x,v) = \yBF(x,v) + \ypf(t,x,v).
\end{equation}
We assume that $\ypf$ is small, so the system can be linearized in $\ypf$. 
Also, because we assume a neutral plasma, the system does not have a background electric field, and
due to the effective external force $F_{\rm ext}$, the background distribution $\yBF(x,v)$ can be inhomogeneous in real space.
Yet, the force $F_{\rm ext}$, which scales as $E^2$, does not enter the final linearized equation.
This leads to the following system of linearized equations:
\begin{subequations}\label{sys:VM-interm}
\begin{eqnarray}
    &&\partial_t \ypf + v\partial_x\ypf - \ypE\partial_v \yBF = 0,\\
    &&\partial_t \ypE - \int v \ypf \diff v = - \yjs.
\end{eqnarray}
\end{subequations}
Here, $v$ is normalized to $\yvth$, $x$ is normalized to the electron Debye length $\lambda_D$, and the time is normalized to the inverse electron plasma frequency $\omegaP^{-1}$, where
\begin{equation}
    \omegaP = \sqrt{\frac{4\pi e^2\ynref }{\yme}}, \quad \yvth = \sqrt{\frac{\yTref}{\yme}}, \quad \lambdaD = \frac{\yvth}{\omegaP},
\end{equation}
where $\ynref$ and $\yTref$ are some fixed values of the electron density and temperature, respectively.
The distribution functions $\ypf$ and $\yBF$ are normalized to  $\ynref/\yvth$, and $\ypE$ is normalized to $\yTref/(e\lambda_D)$. 
We also assume that $\yBF$ is Maxwellian:
\begin{equation}\label{eq:maxwellian}
    \yBF(x,v) =  \frac{n(x)}{\sqrt{2\pi T(x)}} \exp\ylb-\frac{v^2}{2T(x)}\yrb,
\end{equation}
where the background density $n$ and temperature $T$ are normalized to $\ynref$ and $\yTref$, respectively.
An analytical description of the system~\eqref{sys:VM-interm} is presented in Appendix~\ref{app:analytical}.

% ----------------------------------------------------------------------------------------------
% ----------------------------------------------------------------------------------------------
\subsection{Boundary-value problem}\label{sec:bv}
To reformulate Eqs.~\eqref{sys:VM-interm} as a boundary-value problem, we consider a source oscillating at a constant real frequency $\omega_0$:
\begin{equation}
    \yjs(t,x) = \yjs(x) e^{-\yi\omega_0 t}.
\end{equation}
Assuming $\ypf, \ypE \propto \exp(-\yi\omega_0 t)$, one can recast Eqs.~\eqref{sys:VM-interm} as
\begin{subequations}
\begin{eqnarray}
    &&\yi\omega_0 \ypf - v\partial_x \ypf - v H \ypE = 0,\label{eq:vlasov-interm}\\
    &&\yi\omega_0 \ypE + \int v \ypf \diff v = \yjs,
\end{eqnarray}
\end{subequations}
where $H = \yBF/T$ and the variables $g$ and $E$ are now considered the corresponding time-independent complex amplitudes.
For definiteness, we impose a localized current source:
\begin{equation}\label{eq:source-j}
    \yjs = \yi\omega_0 e^{-(x-x_0)^2/(2 \Delta_S^2)}.
\end{equation}
Then, by Eq.~\eqref{eq:charge-conserv}, the corresponding charge density is that of an oscillating dipole:
\begin{equation}\label{eq:charge-source-x}
    \yrhos = -\frac{x-x_0}{\Delta_S^2} e^{-(x-x_0)^2/(2 \Delta_S^2)}.
\end{equation}
We will be interested in the spatial distribution of the electric field $\ypE(x)$ driven by the source $\yjs$ and undergoing linear Landau damping caused by interaction of this field with the distribution perturbation $\ypf$.

To avoid numerical artifacts and keep the grid resolution reasonably low, we impose an artificial diffusivity $\eta$ in the velocity space by modifying Eq.~\eqref{eq:vlasov-interm} as
\begin{equation}\label{eq:vlasov-diff}
    \yi\omega_0 \ypf - v\partial_x \ypf +\eta \partial_v^2\ypf - v H \ypE = 0.
\end{equation} 
This allows us to reduce grids' resolution in both velocity and real space.

To avoid numerical errors caused by the waves reflected from spatial boundaries, we impose outgoing (non-reflecting) boundary conditions on both edges\ycite{Thompson87}.
The resulting system is 
\begin{subequations}\label{sys:boundary-value}
\begin{eqnarray}
    &&\yi\omega_0 \ypf - \zeta^{\rm bc} v\partial_x \ypf +\eta \partial_v^2\ypf - v H \ypE = 0,\label{eq:vlasov-res}\\
    &&\yi\omega_0 \ypE + \int v \ypf \diff v = \yjs,\label{eq:ampere-bva}
\end{eqnarray}
\end{subequations}
where
\begin{equation}\label{eq:bc}
\zeta^{\rm bc} = \left\{ \begin{aligned}
            &0,\quad\text{for incoming waves},\\
            &1,\quad\text{otherwise}.
    \end{aligned} \right. 
\end{equation}
In our case of an one-dimensional system, the incoming waves correspond to $v < 0$ at the right spatial boundary of the simulation box and $v > 0$ at the left spatial boundary.

% ------------------------------------------------------------------------------------
% --- Discretized model ---
% ------------------------------------------------------------------------------------
\section{Discretization}\label{sec:discretization}
To discretize Eqs.~\eqref{sys:boundary-value}, we introduce the following spatial grid:
\begin{equation}\label{eq:grid-x}
    x_j = j h,\quad h = x_{\rm max}/q_x,\quad j = [0, N_x),
\end{equation}
and the following velocity grid:
\begin{equation}\label{eq:grid-v}
    v_k = -v_{\rm max} + k\Delta v,\quad \Delta v = 2v_{\rm max}/q_v,\quad k = [0, N_v).
\end{equation}
Here, $q_x = N_x-1$ and $q_v = N_v-1$, where
\begin{equation}\label{eq:nx-nv}
    N_x = 2^{n_x},\quad N_v = 2^{n_v}.
\end{equation}
For convenience, we also introduce the integer $M_v = 2^{n_v-1}$.
The first $M_v$ points on the velocity grid, $k=[0,M_v)$, correspond to $v_k < 0$, and the last $M_v$ points, $k=[M_v,N_v)$, correspond to $v_k > 0$.
The notation $[k_1,k_2)$, where $k_1$ and $k_2$ are integers, denotes the set of all integers from $k_1$ to $k_2$, including $k_1$ but excluding $k_2$. 
Similarly, the notation $[k_1,k_2]$ denotes the set of integers from $k_1$ to $k_2$, including both $k_1$ and $k_2$.
Also, throughout this paper, the discretized version of any given function $y(x,v)$ is denoted as $y_{j,k}$, where the first subindex is the spatial-grid index and the second subindex is the velocity-grid index

The integral in the velocity space is computed by using the corresponding Riemann sum, $\int y(v)\,\diff v = \sum_k y(v_k)\,\Delta v$. 
To remove $\Delta v$ from discretized equations, we renormalize the distribution functions as follows:
\begin{equation}\label{eq:gF-renorm}
    \Delta v \ypf \to \ypf,\quad \Delta v \yBF\to \yBF.
\end{equation}
In real space, we use the central finite difference scheme:
\begin{subequations}\label{sys:1st-der}
\begin{eqnarray}
    &&\partial_x y_{j,k} = \sigma (y_{j+1,k} - y_{j-1,k}),\label{eq:der-x-bulk}\\
    &&\partial_x y_{0,k} = \sigma (-3y_{0,k} + 4y_{1,k} - y_{2,k}),\label{eq:der-x-left}\\\
    &&\partial_x y_{q_x,k} = \sigma (3 y_{q_x,k} - 4 y_{q_x-1,k} + y_{q_x-2,k}),\label{eq:der-x-right}
\end{eqnarray}
\end{subequations}
where $\sigma = (2h)^{-1}$, $j = [1,N_x-2]$, $k=[0,N_v)$, and the expressions for the derivatives at the boundaries are obtained by considering the Lagrange interpolating polynomial of the second order.
Similarly, the second derivative with respect to velocity is discretized as follows:
\begin{subequations}\label{sys:2nd-der}
\begin{eqnarray}
    &&\partial^2_v y_{j,k} = \beta (y_{j,k+1} - 2 y_{j,k} + y_{j,k-1}),\\
    &&\partial^2_v y_{j,0} = \beta (2 y_{j,0} - 5 y_{j,1}   + 4 y_{j,2} - y_{j,3}),\\
    &&\begin{aligned}
        \partial^2_v y_{j,q_v} &= \\
            &\beta (2 y_{j,q_v} - 5 y_{j,q_v-1} + 4 y_{j,q_v-2} - y_{j,q_v-3}),
    \end{aligned}  
\end{eqnarray}
\end{subequations}
where $\beta = \Delta v^{-2}$, $j = [0,N_x)$, $k=[1,N_v-2]$, and the derivatives at the boundaries are obtained by considering the Lagrange polynomial of the third order.

After the discretization, the Vlasov equation~\eqref{eq:vlasov-res} becomes
\begin{align}\label{eq:vlasov-discretized}
    &P_{j,k} \ypf_{j,k} + \sum_{i=0}^{q_x}P^{(x)}_{j,k,i} g_{i,k} + \sum_{i=0}^{q_v}g_{j,i}P^{(v)}_{i,k} - v_k H_{j,k} \ypE_j = 0,
\end{align}
where $j = [0,N_x)$ and $k=[0,N_v)$.
The function $P_{j,k}$ is given by
\begin{equation}\label{eq:P}
    P_{j,k} = \yi\omega_0 + \zeta^{\rm bc}_{j,k}(\delta_{j,0}-\delta_{j,q_x}) 3 v_k \sigma - p^{\rm sign}_k 2\eta\beta,
\end{equation}
where $\delta_{k_1,k_2}$ is the Kronecker delta:
\begin{equation}
   \delta_{k_1,k_2} = \left\{ \begin{aligned}
                &1,\quad k_1 = k_2,\\
                &0,\quad k_1 \ne k_2.
        \end{aligned} \right.
\end{equation}
The function $p^{\rm sign}_k = 1 - 2(\delta_{k,0}+\delta_{k,q_v})$ appears because the discretization~\eqref{sys:2nd-der} in velocity results in different signs in front of the diagonal element $\ypf_{j,k}$ for bulk and boundary velocity elements.
The coefficient $(\delta_{j,0}-\delta_{j,q_x})$ is necessary to take into account the different signs that appear due to the discretization in space [Eqs.~\eqref{sys:1st-der}] at bulk and boundary points.
The function $\zeta^{\rm bc}_{j,k}$ is the discretized version of the function~\eqref{eq:bc} responsible for the outgoing boundary conditions:
\begin{equation}
   \zeta^{\rm bc}_{j,k} = \left\{ \begin{aligned}
                &0,\quad j = 0,\ \ \ k = [M_v, N_v),\\
                &0,\quad j = q_x,\   k = [0, M_v),\\
                &1,\quad\text{otherwise}.
        \end{aligned} \right.
\end{equation}
The function $P^{(x)}_{j,k,i}$ varies for bulk and boundary spatial points according to Eqs.~\eqref{sys:1st-der}:
\begin{equation}\label{eq:Px}
    P^{(x)}_{j,k,i} = v_k \sigma \zeta^{\rm bc}_{j,k} \times 
        \left\{ \begin{aligned}
                &\delta_{j-1, i} - \delta_{j+1, i}, \quad\ \ j = [1,q_x),\\
                &\delta_{j+2, i} - 4\delta_{j+1, i},\quad   j = 0,\\
                &4\delta_{j-1, i} - \delta_{j-2, i},\quad   j = q_x.
        \end{aligned} \right.
\end{equation}
The function $P^{(v)}_{i,k}$ varies for bulk and boundary velocity points according to Eqs.~\eqref{sys:2nd-der}:
\begin{equation}\label{eq:Pv}
    P^{(v)}_{i,k} = -\eta\beta \times 
        \left\{ \begin{aligned}
                &-(\delta_{i,k+1} + \delta_{i,k-1}),  \quad\quad\quad\ \ k = [1, q_v),\\
                & 5\delta_{i,k+1} - 4\delta_{i,k+2} + \delta_{i,k+3},\quad k = 0,\\
                & 5\delta_{i,k-1} - 4\delta_{i,k-2} + \delta_{i,k-3},\quad k = q_v.
        \end{aligned} \right.
\end{equation}
Finally, Amp\`ere law~\eqref{eq:ampere-bva} is recasted as
\begin{equation}\label{eq:ampere-discretized}
    \yi\omega_0 \ypE_j + \sum_{k=0}^{N_v-1} v_k \ypf_{j,k} = \yjs_j,
\end{equation}
where $j=[0,N_x)$.

\section{Matrix representation}\label{sec:linear-equations}

After the discretization, Eqs.~\eqref{eq:vlasov-discretized} and~\eqref{eq:ampere-discretized} can be converted into the form of Eq.~\eqref{eq:axb} as follows.

% ------------------------------------------------------------------------------------
% ------------------------------------------------------------------------------------
\subsection{The vector $\mathbf{\ypsi}$}\label{sec:encoding-psi}
First of all, to construct Eq.~\eqref{eq:axb}, one needs to encode $\ypf_{j,k}$ and $E_j$ into $\ypsi$.
Since one needs to store $N_f = 2$ fields on a $N_x\times N_v$ phase space, the size of $\ypsi$ should be $N_{\rm tot} = N_f N_{xv}$ where $N_{xv} = N_x N_v$, and this vector can be saved by using $1+n_x+n_v$ qubits, where $n_x$ and $n_v$ have been introduced in Eq.~\eqref{eq:nx-nv}.
Within the vector $\ypsi$, the fields are arranged in the following way:
\begin{equation}\label{eq:psi}
\psi_{d N_{xv} + j N_v + k} = 
    \left\{ \begin{aligned}
        \ypf_{j,k}&,\quad d = 0,\\
        \delta_{k,0} \ypE_j&,\quad d = 1,\\
        0&,\quad\text{otherwise},
    \end{aligned}\right.
\end{equation}
where $\psi_{d N_{xv} + j N_v + k}$ are the elements of $\ypsi$ with $j = [0, N_x)$, $k = [0,N_v)$, and $d = [0, N_f)$.
Since the electric field does not depend on velocity, the second half of $\ypsi$ is filled with $E_j$ only if the velocity index $k$ is equal to zero. 

To address each field, $\ypf_{j,k}$ or $E_j$, we introduce the register $r_f$ with one qubit. 
The zero state $\ket{0}_{r_f}$ corresponds to addressing the plasma distribution function, and the unit state $\ket{1}_{r_f}$ flags the electric field.
We also use additional two registers, denoted $r_x$ and $r_v$, with $n_x$ and $n_v$ qubits, respectively, to specify the fields' position in the real and velocity spaces, correspondingly.
Then, one can express the vector $\ypsi$ as follows:
\begin{equation}\label{eq:psi-ket}
\begin{split}
\ket{\psi} = &\eta_{\psi, {\rm norm}} \sum_{j=0}^{N_x - 1}\bigg(\sum_{k=0}^{N_v-1} \ypf_{j,k}\ket{0}_{r_f}\ket{j}_{r_x}\ket{k}_{r_v}\\ 
    & + \ypE_j\ket{1}_{r_f}\ket{j}_{r_x}\ket{0}_{r_v}\bigg),
\end{split}
\end{equation}
where $\eta_{\psi, {\rm norm}}$ is the normalization factor used to ensure that $\itimes{\psi}{\psi} = 1$.
(The assumed notation is such that the least significant qubit is the rightmost qubit. In quantum circuits, the least significant qubit is the lowest one.)

% ------------------------------------------------------------------------------------
% ------------------------------------------------------------------------------------
\subsection{The source term}\label{sec:encoding-source}
Corresponding to Eq.~\eqref{eq:psi}, the discretized version of the vector $\yyb$ in Eq.~\eqref{eq:axb} is as follows:
\begin{equation}\label{eq:encoding-b}
b_{d N_{xv} + j N_v + k} = 
    \left\{ \begin{aligned}
        \delta_{k,0}\yjs_j&,\quad d = 1,\\
        0&,\quad d = 0.
    \end{aligned}\right.
\end{equation}
Note that all elements in the first half of $\yyb$ are zero, and, in its second half, only every $N_v$-th element is nonzero.
In the `ket' notation, this can be written as
\begin{equation}\label{eq:b}
    \ket{b} = \eta_{b,{\rm norm}}\sum_{j = 0}^{N_x-1} \yjs_j\ket{1}_{r_f}\ket{j}_{r_x}\ket{0}_{r_v},
\end{equation}
where $\eta_{b,{\rm norm}}$ is the normalization factor used to ensure that $\itimes{b}{b} = 1$.

% ------------------------------------------------------------------------------------
% ------------------------------------------------------------------------------------
\subsection{The matrix $\yA$}\label{sec:matrix}

% -------------------------------------------------------------------------------------------------------
% -------------------------------------------------------------------------------------------------------
\begin{figure}[!t]
\centering
\includegraphics[width=0.48\textwidth]{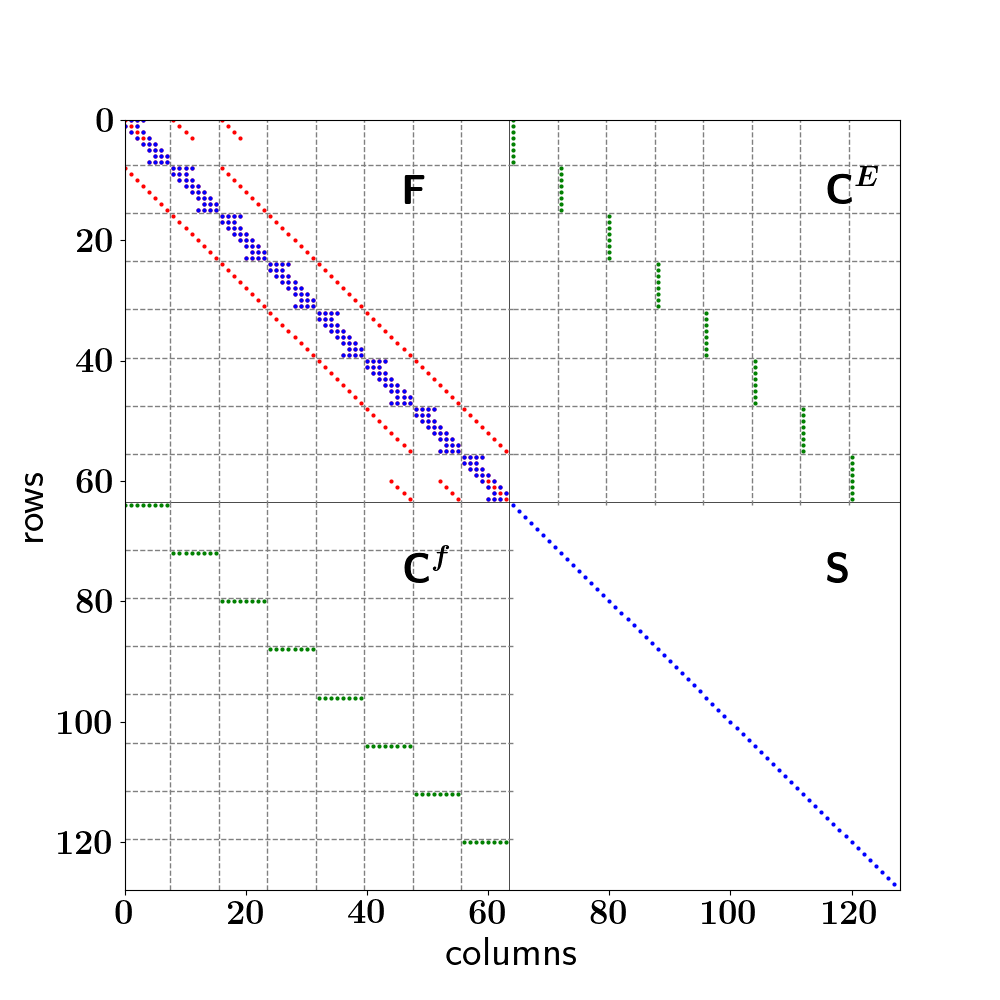}
\caption{\label{fig:A-colored-structure} 
    A schematic showing the structure of the matrix $\yA$ [Eq.~\eqref{eq:A}] with $n_x = 3$ and $n_v = 3$.
    The solid lines separate the submatrices introduced in Eq.~\eqref{eq:A}.
    The dashed lines indicate blocks of size $N_v\times N_v$.
    The blue markers indicate velocity-independent elements.  
}\end{figure}
% -------------------------------------------------------------------------------------------------------
% -------------------------------------------------------------------------------------------------------

The corresponding $N_{\rm tot}\times N_{\rm tot}$ matrix $\yA$ is represented as 
\begin{equation}\label{eq:A}
\yA = 
\begin{pmatrix}
    \yF   & \yC^E \\
    \yC^f & \yS
\end{pmatrix},
\end{equation}
where $\yF$, $\yC^E$, $\yC^f$, and $\yS$ are $N_{xv}\times N_{xv}$ submatrices. 
A schematic structure of this matrix is shown in Fig.~\ref{fig:A-colored-structure}.
The submatrix $\yF$ encodes the coefficients in front of $\ypf$ in the Vlasov equation~\eqref{eq:vlasov-discretized} and is given by
\begin{equation}\label{eq:F}
\yF = 
\begin{pmatrix}
    \yF^{\rm L,0}        &\yF^{\rm L,1}  & \yF^{\rm L,2} & 0                      & 0                      &\dots                   \\
    \yF^{\rm B,1} &\yF^{\rm B,0}            &-\yF^{\rm B,1}     & 0                      & 0                      &\dots                   \\
    0                  &\yF^{\rm B,1}     & \yF^{\rm B,0}            &-\yF^{\rm B,1}     & 0                      &\dots                   \\
    \dots              &\dots                  & \dots                  & \dots                  & \dots                  &\dots                   \\
    \dots              & 0                     & \yF^{\rm B,1}     & \yF^{\rm B,0}            &-\yF^{\rm B,1}    & 0                      \\
    \dots              & 0                     &             0          & \yF^{\rm B,1}     & \yF^{\rm B,0}            &-\yF^{\rm B,1}     \\
    \dots              & 0                     &             0          & \yF^{\rm R,2} & \yF^{\rm R,1} & \yF^{\rm R,0}     
\end{pmatrix}.
\end{equation}
This submatrix consists of $N_x^2$ blocks, whose elements are mostly zeros.
Each block is of size $N_v\times N_v$, and the position of each block's elements is determined by the row index $k_r = [0, N_v)$ and the column index $k_c = [0, N_v)$.
The discretization at the left spatial boundary is described by the blocks $\yF^{\rm L,0}$, $\yF^{\rm L,1}$, and $\yF^{\rm L,2}$.
The discretization at the right spatial boundary is described by the blocks $\yF^{\rm R,0}$, $\yF^{\rm R,1}$, and $\yF^{\rm R,2}$.
The discretization at the bulk spatial points is described by the blocks $\yF^{\rm B,0}$ and $\yF^{\rm B,1}$.
Using Eqs.~\eqref{eq:P},~\eqref{eq:Px}, and ~\eqref{eq:Pv}, one obtains the following expressions for the elements in each block in~\eqref{eq:F} at the left spatial boundary:
\begin{subequations}\label{eq:FL}
\begin{eqnarray}
    &&F^{\rm L,0}_{k_r, k_c} = \delta_{k_r, k_c} P_{0,k_r} + P^{(v)}_{k_c, k_r},\label{eq:FL0}\\
    &&F_{k_r, k_c}^{\rm L,1} = -4 v_{k_r} \sigma \zeta^{\rm bc}_{0,k_r} \delta_{k_r, k_c},\\
    &&F_{k_r, k_c}^{\rm L,2} = v_{k_r} \sigma \zeta^{\rm bc}_{0,k_r} \delta_{k_r, k_c},
\end{eqnarray}
\end{subequations}
at the right spatial boundary:
\begin{subequations}\label{eq:FR}
\begin{eqnarray}
    &&F^{\rm R,0}_{k_r, k_c} = \delta_{k_r, k_c} P_{q_x,k_r} + P^{(v)}_{k_c, k_r},\label{eq:FR0}\\
    &&F_{k_r, k_c}^{\rm R,1} = 4 v_{k_r} \sigma \zeta^{\rm bc}_{q_x,k_r} \delta_{k_r, k_c},\\
    &&F_{k_r, k_c}^{\rm R,2} = -v_{k_r} \sigma \zeta^{\rm bc}_{q_x,k_r} \delta_{k_r, k_c}.
\end{eqnarray}
\end{subequations}
and at bulk spatial points:
\begin{subequations}\label{eq:FB}
\begin{eqnarray}
    &&F^{\rm B,0}_{k_r, k_c} = \delta_{k_r, k_c} P_{j,k_r} + P^{(v)}_{k_c, k_r},\\
    &&F_{k_r, k_c}^{\rm B,1} = v_{k_r} \sigma \delta_{k_r, k_c}.\label{eq:FB1}
\end{eqnarray}
\end{subequations}
We denote the part of the submatrix $\yF$ that depends on $v$ as $\yFP$.
The matrix elements of $\yFP$ are indicated in Fig.~\ref{fig:A-colored-structure} with red.
The part of $\yF$ which does not depend on $v$ is denoted as $\yFF$ and shown in blue in Fig.~\ref{fig:A-colored-structure} within the submatrix $\yF$.

The matrix $\yC^E$ is block-diagonal and encodes the coefficients in front of $\ypE$ in the Vlasov equation~\eqref{eq:vlasov-discretized}:
\begin{equation}\label{eq:CE}
    C^E_{j_r N_v + k_r, j_c N_v + k_c} = 
        -\delta_{j_r,j_c}\delta_{k_c,0}v_{k_r} H_{j_r,k_r},
\end{equation}
where $j_r,j_c = [0, N_x)$ and $k_r, k_c = [0,N_v)$.
In this submatrix, the first column in each diagonal block of size $N_v\times N_v$ is nonsparse whilst all other columns are filled with zeroes.
The Kronecker delta $\delta_{k_c,0}$ appears because of the chosen encoding of the electric field into the state vector $\ypsi$ according to Eq.~\eqref{eq:psi}.

The submatrix $\yC^f$ is also block-diagonal and encodes the coefficients in front of $\ypf$ in Eq.~\eqref{eq:ampere-discretized}:
\begin{equation}\label{eq:Cf}
    C^f_{j_r N_v + k_r, j_c N_v + k_c} = \delta_{j_r,j_c}\delta_{k_r,0}v_{k_c},
\end{equation}
where $j_r,j_c = [0, N_x)$ and $k_r, k_c = [0,N_v)$.
Here, the first row in each block of size $N_v\times N_v$ is nonsparse due to the sum in Eq.~\eqref{eq:ampere-discretized}.

Finally, the matrix $\yS$ is diagonal and encodes the coefficients in front of $\ypE$ in Eq.~\eqref{eq:ampere-discretized}:
\begin{equation}\label{eq:S}
S_{j_r N_v + k_r, j_c N_v + k_c} = 
    \delta_{j_r,j_c}\delta_{k_r,k_c}\yi\omega_0,
\end{equation}
where $j_r,j_c = [0, N_x)$ and $k_r, k_c = [0,N_v)$.
% Here, we keep $\yi\omega_0$ for $k_r > 0$ to ensure that the resulting matrix $\yA$ is invertible.

% -------------------------------------------------------------------------------------------------------
% -------------------------------------------------------------------------------------------------------
\begin{figure}[t]
\centering
\hspace{-1cm}
\includegraphics[width=0.48\textwidth]{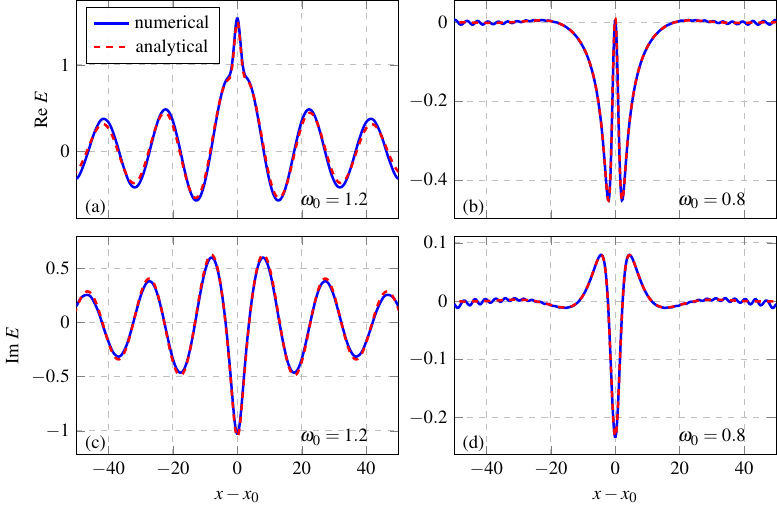}
\caption{
    \label{fig:E-num-analytics-wo-diff} 
    Plots showing the spatial distribution of the electric field computed numerically (blue) and analytically (red) using Eq.~\eqref{eq:E-analytical}. 
    (a), (c): ${\rm Re}\,E$ and ${\rm Im}\,E$, respectively, for $\omega_0 = 1.20$.
    One can see Langmuir waves propagating away from the source (located at $x = x_0$) and experiencing weak Landau damping.
    (b), (d): ${\rm Re}\,E$ and ${\rm Im}\,E$, respectively, for $\omega_0 = 0.8$.
    One can see Debye shielding of the source charge.
    In both cases, $n_x = 9$, $n_v = 8$, and $\eta = 0$. 
}\end{figure}
% -------------------------------------------------------------------------------------------------------
% -------------------------------------------------------------------------------------------------------
\begin{figure}[!t]
\centering
\subfloat{\includegraphics[width=0.48\textwidth]{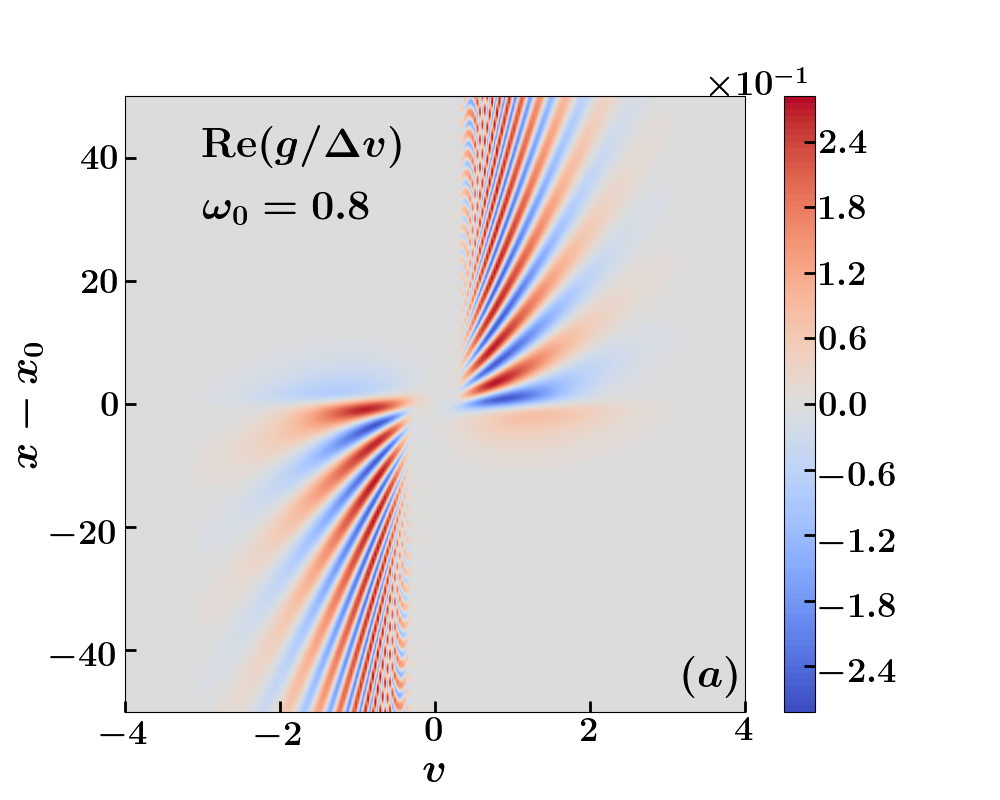}}\\
\vspace{-0.8cm}
\subfloat{\includegraphics[width=0.48\textwidth]{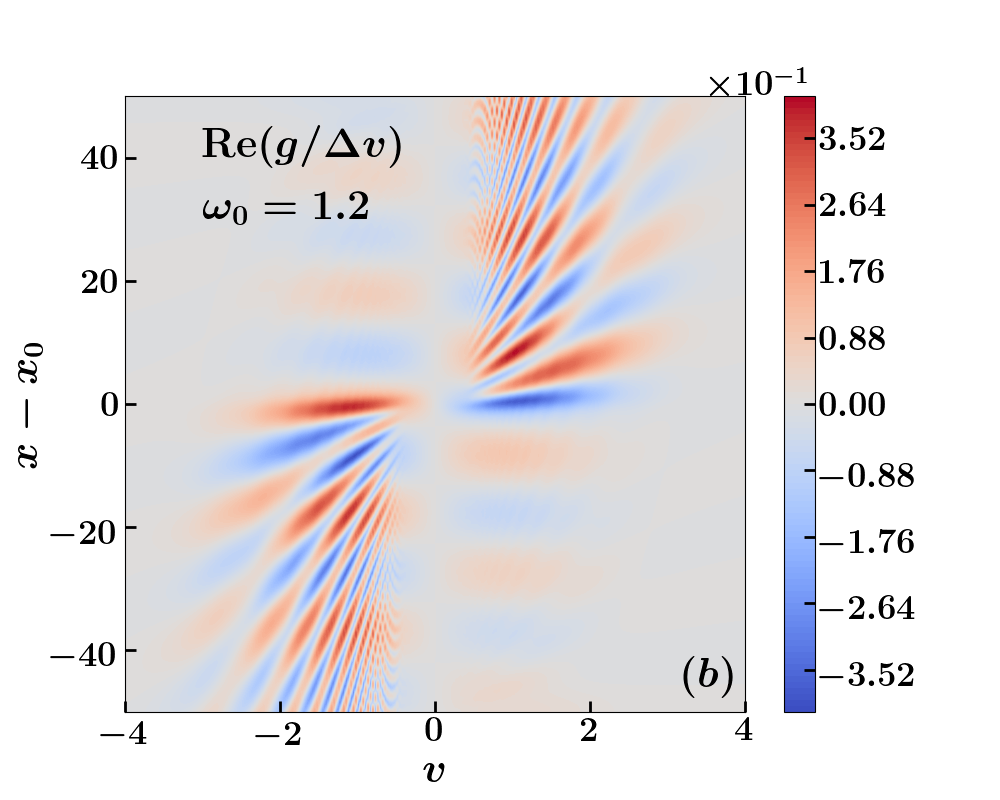}}
\caption{
    \label{fig:f-woD} 
    Plots showing the real component of the plasma distribution function, in units $\Delta v$, computed numerically with $n_x = 9$, $n_v = 8$, and $\eta = 0.0$. 
    (a): $\omega_0 = 0.8$. (b):  $\omega_0 = 1.2$.
    % Here, the plasma distribution is divided by $\Delta v$ because of Eq.~\eqref{eq:gF-renorm}.
}
\end{figure}
% -------------------------------------------------------------------------------------------------------
% -------------------------------------------------------------------------------------------------------

\section{Classical simulations}\label{sec:classical-results}

% -------------------------------------------------------------------------------------------------------
% -------------------------------------------------------------------------------------------------------
\begin{figure}[t]
\centering
\hspace{-1cm}
\includegraphics[width=0.48\textwidth]{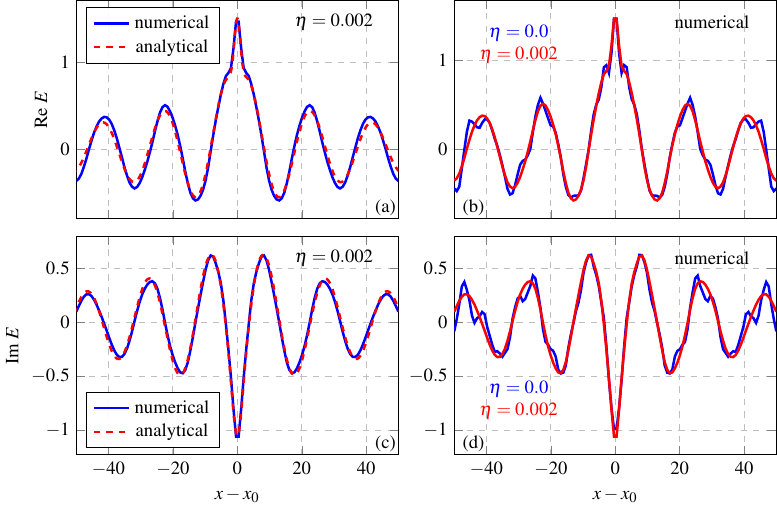}
\caption{
    \label{fig:E-num-analytics-with-diff} 
    Plots showing the spatial distribution of the electric field for $\omega_0 = 1.2$, $n_x = 7$, and $n_v = 5$.
    (a), (c): Results from the numerical (blue) and analytical (red) computations with the diffusivity $\eta = 0.002$. 
    (b), (d): Results from the numerical computations with (red) and without (blue) diffusivity. 
}\end{figure}
% -------------------------------------------------------------------------------------------------------
% -------------------------------------------------------------------------------------------------------
\begin{figure}[!t]
\centering
\subfloat{\includegraphics[width=0.48\textwidth]{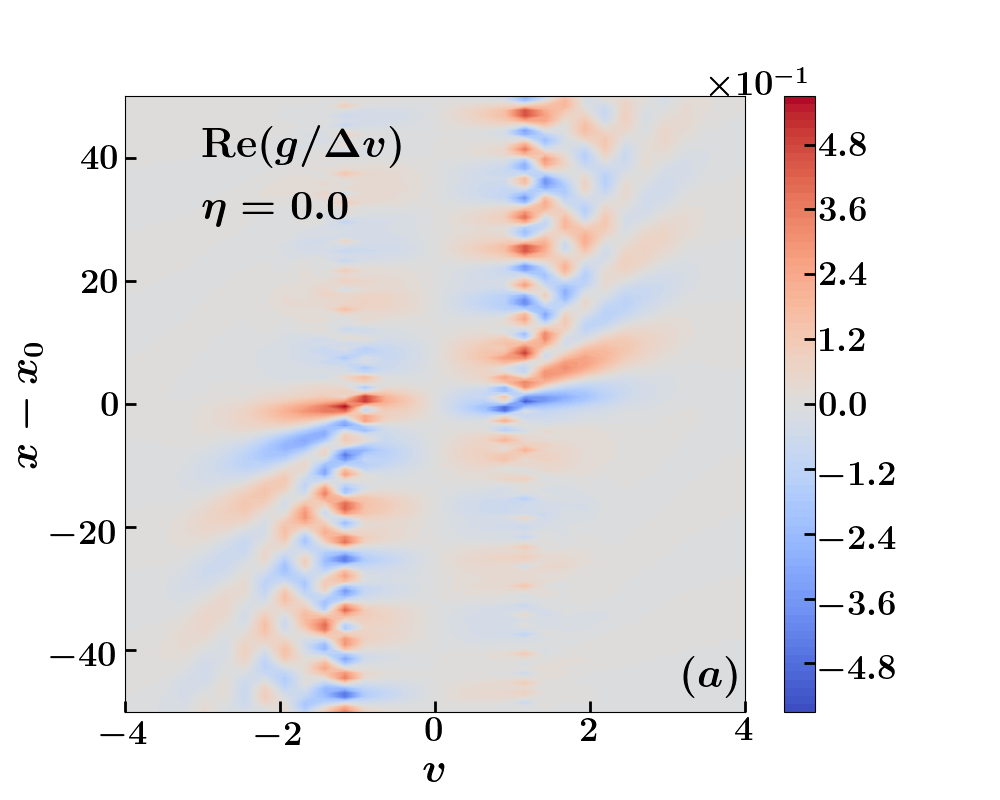}}\\
\vspace{-0.8cm}
\subfloat{\includegraphics[width=0.48\textwidth]{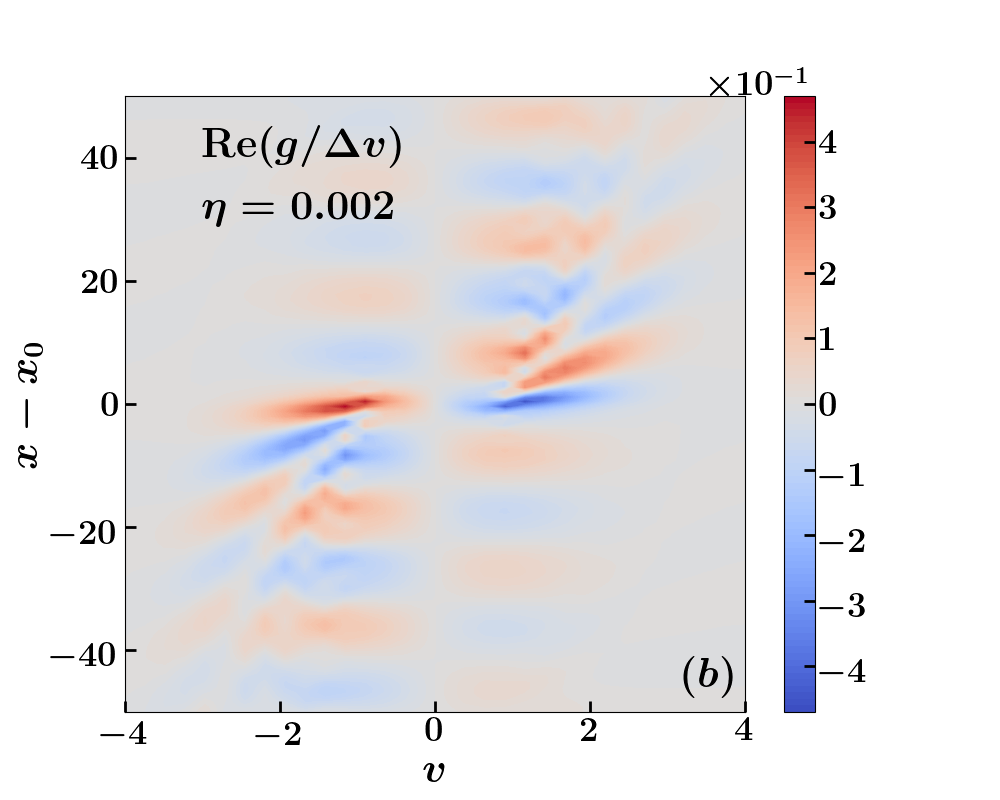}}
\caption{
    \label{fig:f-w12-D} 
    Plots showing the real component of the plasma distribution function, in units $\Delta v$, computed numerically for the cases with $\omega_0=1.2, n_x = 7$, and $n_v = 5$. 
    (a): $\eta = 0.0$. (b):  $\eta = 0.002$.
    % Here, the plasma distribution is divided by $\Delta v$ because of Eq.~\eqref{eq:gF-renorm}.
}
\end{figure}
% -------------------------------------------------------------------------------------------------------
% -------------------------------------------------------------------------------------------------------

To test our discretization scheme, we have performed classical simulations for homogeneous plasma ($n = T = 1$), which facilitates comparison with the analytic theory described in Appendix~\ref{app:analytical}. 
In our simulations, the phase space is described by $x_{\rm max} = 100, n_x = 9$ and $v_{\rm max} = 4, n_v = 8$.
The source in the form~\eqref{eq:source-j} is placed at $x_0 = 50$ with $\Delta_S = 1.0$.
We consider two cases (Fig.~\ref{fig:E-num-analytics-wo-diff}): 
(a) $\omega_0 = 1.2$, which corresponds to the case when the source frequency exceeds the plasma frequency, and 
(b) $\omega_0 = 0.8$, which corresponds to the case when the plasma frequency exceeds the source frequency.
The numerical calculations were performed 
% by storing the matrix~\eqref{eq:A} in a sparse form (only nonzero elements were considered) with its subsequent inversion 
by inverting the matrix~\eqref{eq:A} using the sparse-QR-factorization-based method provided in CUDA toolkit cuSOLVER\ycite{EVMCL, cuSolver}.

In the case with $\omega_0 = 1.2$, the source launches Langmuir waves propagating outward and gradually dissipating via Landau damping. 
The outgoing boundary conditions allow the propagating wave leave the simulated box with negligible reflection.
In the case with $\omega_0 = 0.8$, the plasma shields the electric field, which penetrates plasma roughly by a Debye length.
Due to the high resolution in both real and velocity space, the model remains stable and does not generate visible numerical artifacts (Fig.~\ref{fig:f-woD}).
Artifacts become noticeable at lower resolution (Figs.~\ref{fig:E-num-analytics-with-diff} and~\ref{fig:f-w12-D}) but can be suppressed by introducing artificial diffusivity $\eta$ in the velocity space [Eq.~\eqref{eq:vlasov-diff}]. 
Such simulations are demonstrated in Figs.~\ref{fig:E-num-analytics-with-diff} and~\ref{fig:f-w12-D} for $n_x = 7, n_v = 5$, and $\eta = 0.002$.
As seen in Fig.~\ref{fig:E-num-analytics-with-diff}, their results are in good agreement with the analytical solution.
However, keep in mind that introducing diffusivity complicates BE and somewhat increases the condition number of $\yA$. 
For example, if one takes $n_x = 7$, $n_v = 5$, $\eta = 0.002$, $\omega = 1.2$, the condition number of the resulting matrix is $\kappa_A = 8.844\times 10^4$ (i.e. $\log_2\kappa_A = 16.4$).
Without the diffusivity and with the same resolution, the condition number is $\kappa_A = 3.489\times 10^4$ (i.e. $\log_2\kappa_A = 15.1$).

\section{Encoding the equations into a quantum circuit}\label{sec:encoding}

% ----------------------------------------------------------------------------------------
% --- Preparation ---
% ----------------------------------------------------------------------------------------
\subsection{Initialization}\label{sec:initialization}
To encode the right-hand-side vector~\eqref{eq:b} into a quantum circuit, one can use the fact that the shape of the source current~\eqref{eq:source-j} is Gaussian.
As shown in Refs.~\yocite{Novikau23, Kane23, Hariprakash23}, one can encode this function by using either QSVT\ycite{Gilyen19, Martyn21} or the so-called Quantum Eigenvalue Transformation of Unitaries (QETU)\ycite{Dong22}, where the scaling of the resulting circuit is $\oO(n_v \log_2(\varepsilon^{-1}_{\rm qsvt}))$ and $\varepsilon_{\rm qsvt}$ is the desired absolute error in the QSVT approximation of the Gaussian.
However, one should keep in mind that the success probability of the initialization circuit depends on the Gaussian width.
To increase this probability, amplitude amplification can be used\ycite{Brassard02}.

% ----------------------------------------------------------------------------------------
% --- Preparation ---
% ----------------------------------------------------------------------------------------
\subsection{Block encoding: basic idea}

% ---------------------------------------------------------
\begin{figure*}[!t]
\centering
\includegraphics[width=0.96\textwidth]{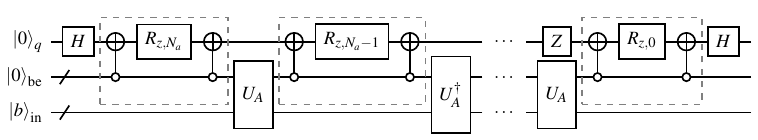}
\caption{
    \label{circ:qsvt} 
    The QSVT circuit encoding a real polynomial of order $N_a$, where $N_a$ is odd, by using $N_a + 1$ angles $\phi_k$ pre-computed classically.
    The gates denoted as $R_{z,k}$ represent the rotations $R_z(2\phi_k)$.
    For an even $N_a$, the gate $Z$ should be removed and the rightmost BE oracle $U_A$ should be replaced with its Hermitian adjoint version $U_A^\dagger$. 
}
\end{figure*}
% ---------------------------------------------------------

To solve Eq.~\eqref{eq:axb} with the matrix~\eqref{eq:A} and the source~\eqref{eq:b} on a digital quantum computer, one can use the QSVT, which approximates the inverse matrix $\yA^{-1}$ with an odd polynomial of the matrix singular values.
(The mathematical foundations of the QSVT are described, for example, in Refs.~\yocite{Gilyen19, Martyn21, Lin22}.)
The QSVT returns a quantum state $\ket{\psi_{\rm qsvt}}$ whose projection on zero ancillae is proportional to the solution $\ypsi$ of Eq.~\eqref{eq:axb}:
\begin{equation}
    \ket{\psi_{\rm qsvt}} = \frac{e^{\yi\phi_{\rm glob}}}{\kappa_{\rm qsvt}}\yA^{-1}\ket{0}_{\rm anc}\ket{b}_{\rm in} + \ket{\neq 0}_{\rm anc}\ket{\ldots}_{\rm in},
\end{equation}
where $\phi_{\rm glob}$ is an unknown global angle and the scalar parameter $\kappa_{\rm qsvt}$ is of the order of the condition number of the matrix $\yA$.

A typical QSVT circuit is shown in Fig.~\ref{circ:qsvt}, where the angles $\phi_i$ are computed classically.
These angles serve as the parameters that specify the function computed by the QSVT circuit. 
In our case, the function is the inverse function. 
(More details about the computation of the QSVT angles can be found in  Refs.~\yocite{Dong21, Ying22}).
The subcircuit $U_A$ is the so-called block-encoding (BE) oracle with the following matrix representation:
\begin{equation}\label{eq:UA}
\yU_A = 
    \begin{pmatrix}
        \yA & \cdot \\
        \cdot & \cdot
    \end{pmatrix},
\end{equation}
where $\yU_A$ is a unitary matrix.
(The dots correspond to submatrices that keep $\yU_A$ unitary but otherwise are unimportant.)
This unitary encodes the matrix $\yA$ as a subblock that is accessed by setting the ancilla register `be' to zero:
\begin{equation}
    \yU_A\ket{0}_{\rm be}\ket{\psi}_{\rm in} = \yA\ket{0}_{\rm be}\ket{\psi}_{\rm in} + \ket{\neq 0}_{\rm be}\ket{\ldots}_{\rm in}.
\end{equation}
The QSVT addresses the BE oracle $\oO(\kappa_{\rm qsvt}\log_2(\varepsilon^{-1}_{\rm qsvt}))$ times to approximate the inverse matrix $\yA^{-1}$.
The efficient implementation of the BE oracle is the key to a potential quantum speedup that might be provided by the QSVT.
As discussed in Ref.~\yocite{Novikau23}, the QSVT can provide a polynomial speedup for two- and higher-dimensional classical wave systems if the quantum circuit of the BE oracle scales polylogarithmically or better with the size of $\yA$. 

% To block encode a Hermitian matrix, one can use the technique proposed in Ref.~\yocite{Berry12} and applied, for instance, in Ref.~\yocite{Novikau22} for QSP modeling of linear waves in fluid plasmas.
To block encode a non-Hermitian matrix such as $\yA$, one can first extend it to a Hermitian one as
\begin{equation}\label{eq:A-ext}
    \yA_{\rm ext} = 
    \begin{pmatrix}
        0         & \yA \\
        \yA^\dagger & 0
    \end{pmatrix}
\end{equation}
and then use the technique from Ref.~\yocite{Berry12}.
However, this will require at least two additional ancilla qubits for the extension~\eqref{eq:A-ext}.
Another option is to decompose $\yA$ into two Hermitian matrices:
\begin{equation}\label{eq:A-as-sum}
    \yA = \yA_h + \yi \yA_a,
\end{equation}
where $\yA_h = (\yA + \yA^\dagger)/2$ and $\yA_a = (\yA - \yA^\dagger)/(2\yi)$.
The sum~\eqref{eq:A-as-sum} can be computed by using the circuit of Linear Combination of Unitaries (LCU), which requires an additional ancilla.
Also note that the matrices $\yA_a$ and $\yA_h$, although being Hermitian, still have non trivial structure.
Thus, in this method, it is necessary to block encode two separate matrices, which may double the depth of the BE oracle.

To reduce the number of ancillae and avoid encoding two matrices instead of one, we 
propose to encode the non-Hermitian $\yA$ directly, without invoking the extension~\eqref{eq:A-ext} or the splitting~\eqref{eq:A-as-sum}. 
This technique was already used in Ref.~\yocite{Novikau23}.
Although the direct encoding requires ad hoc construction of some parts of the BE oracle, this approach leads to a more compact BE circuit.
 
To encode $\yA$ into the unitary $\yU_A$, one needs to normalize $\yA$ such that $\varsigma||\yA||_{\rm max} \leq 1$, where
\begin{equation}
    ||\yA||_{\rm max} = \max_k\sum_j\sqrt{|A_{kj}|^2}
\end{equation}
and $\varsigma$ is related to the matrix nonsparsity as will be explained later [see Eq.~\eqref{eq:nonsparsity}].
Here, by the term `nonsparsity', we understand the maximum number of nonzero elements in a matrix row or column.
\footnote{Although the term `sparsity' is commonly used in literature, it is better to use the term ``nonsparsity'' when one refers to the parameter $\varsigma$ which grows than a matrix becomes less sparse.
In other words, matrices with a large $\varsigma$ are characterized by low sparsity.}
Hence, $\yA$ should be normalized as follows:
\begin{equation}\label{eq:first-normalization-A}
    \yA \to \yA/\ylb||\yA||_{\rm max}\varsigma\yrb.
\end{equation}

The general structure of a BE oracle implementing the direct encoding of a non-Hermitian matrix is
\begin{equation}\label{eq:UA-general}
    U_A = O^\dagger_F O^{\rm B}_{F,{\rm corr}} O_M O_H O^{\rm F}_{F,{\rm corr}} O_F,
\end{equation}
where the oracles $O_F$, $O^{\rm F}_{F,{\rm corr}}$, $O_M$, $O^{\rm B}_{F,{\rm corr}}$, and $O^\dagger_F$ encode the positions of nonzero matrix elements in $\yA$, and $O_H$ encodes the values of these elements.
The upper superscripts ``F'' and ``B''  stand for ``forward'' and ``backward'' action.
The oracle $U_A$ can be constructed by using quantum gates acting on a single qubit but controlled by multiple qubits.
Here, such gates are called single-target multicontrolled (STMC) gates (Sec.~\ref{sec:STMC}).

Equation~\eqref{eq:UA-general} is based on the BE technique from Ref.~\yocite{Berry12} with the only difference that the oracles $O^{\rm F}_{F,{\rm corr}}$ and $O^{\rm B}_{F,{\rm corr}}$ are introduced to take into account the non-Hermiticity of $\yA$ and are computed ad hoc by varying STMC gates.
Basically, the oracles $O_F$, $O_M$, and $O^\dagger_F$ create a structure of a preliminary Hermitian matrix which is as close as possible to the target non-Hermitian matrix $\yA$.
The structure is then corrected by the oracles $O^{\rm F}_{F,{\rm corr}}$ and $O^{\rm B}_{F,{\rm corr}}$ constructed by varying their circuits to encode the structure of $\yA$.

We consider separately the action of the following oracle:
\begin{equation}\label{eq:UD}
    U_D = O^\dagger_F O^{\rm B}_{F,{\rm corr}} O_M O^{\rm F}_{F,{\rm corr}} O_F,
\end{equation}
where the oracle $O_H$ is not included.
The matrix representation of the oracle $U_D$ projected onto zero ancillae is denoted as $\yD^A$.
The matrix $\yD^A$ has nonzero elements at the same positions as the ones of the nonzero elements of $\yA$.

We use the input register `in' to encode a row index $i_r$ and the ancilla register $a_c$ to perform intermediate computations.
Then,
\begin{equation}\label{eq:UD-action}
    U_D\ket{0}_{a_c}\ket{i_r}_{\rm in} = \sum_{i_c = 0}^{N_{c,i_r}-1} d_{i_r i_{c}} \ket{0}_{a_c}\ket{i_{c}}_{\rm in} + \ket{\neq 0}_{a_c}\ket{\ldots}_{\rm in},
\end{equation}
where $i_{c}$ are the column indices of all nonzero elements of $\yA$ at the row $i_r$; $N_{c,i_r}$ is the number of these elements at $i_r$; $d_{i_r i_{c}} \leq 1$ are the matrix elements of $\yD^A$.
The number $N_{c,i_r}$ is less or equal to the nonsparsity of $\yA$ and can be different at different $i_c$.
The elements $d_{i_r i_{c}}$ are usually different powers of the factor $2^{-1/2}$ which appears due to the usage of  multiple Hadamard gates $H$.

The oracle $O_H$, which enters $U_A$ but is not a part of $U_D$, takes the matrix $\yD^A$ and modifies its elements to form $\yA$. Usually, $O_H$ acts on an extra ancilla $a_e$ which is not used by $U_D$.
Due to that, we can formally consider $U_D$ separately from $O_H$.
The action of $O_H$ can be considered as the following mapping:
\begin{equation}\label{eq:OH-mapping}
    O_H: d_{i_r i_c} \to A_{i_r i_c},
\end{equation}
where $A_{i_r i_c}$ are elements of the matrix $\yA$ after the normalization~\eqref{eq:first-normalization-A}.
For instance, to encode a real-value element $A_{i_r i_c}$, one can use the rotation gate $R_y(\theta)$:
\begin{equation}
    R_y(\theta) d_{i_r i_c}\ket{0}_{a_e} = \cos(\theta/2)d_{i_r i_c}\ket{0}_{a_e} + \ldots\ket{1}_{a_e}.
\end{equation}
The factor $d_{i_r i_c}$ appears from the action of the oracle $U_D$ [Eq.~\eqref{eq:UD-action}].
Our goal is to have $A_{i_r i_c} = \cos(\theta/2)d_{i_r i_c}$.
Thus,
\begin{equation}\label{eq:OH-action}
    \theta = 2 \arccos(A_{i_r i_c} / d_{i_r i_c}).
\end{equation}
The fact that $d_{i_r i_c} \leq 1$ is the reason why it is necessary to include $\varsigma$ into the normalization~\eqref{eq:first-normalization-A}.
From this, we conclude that 
\begin{equation}\label{eq:nonsparsity}
    \varsigma = {\rm max}_{ij}|(D^A_{ij})^{-1}|.
\end{equation}

The oracle $O_H$ usually consists of STMC rotations gates $R_x$, $R_y$, $R_z$, and $R_c$ [Eq.~\eqref{eq:Rc}].
The first two are used to encode imaginary and real values, correspondingly.
The third one can be used to change the sign of a value or to turn a real value into an imaginary one if necessary, and vice versa.
The gate $R_c$ is used to encode complex values.

Now, let us specify the action of different parts of $U_A$.
The oracles $O^{\rm F}_{F,{\rm corr}} O_F$ encode column indices into the ancilla register $a_c$:
\begin{equation}\label{eq:OFF}
    O^{\rm F}_{F,{\rm corr}} O_F\ket{0}_{a_e}\ket{0}_{a_c}\ket{i_r}_{\rm in} = \sum_{i_c}\sqrt{d_{i_r i_c}}\ket{0}_{a_e}\ket{i_c}_{a_c}\ket{i_r}_{\rm in}.
\end{equation}
The oracle $O_H$ uses the row index from the state register `in' and the column indices from the ancilla register $a_c$ to determine which element should be computed and then encodes it into the state amplitude:
\begin{equation}
\begin{split}
    O_H \sum_{i_c}&\sqrt{d_{i_r i_c}}\ket{0}_{a_e}\ket{i_c}_{a_c}\ket{i_r}_{\rm in}\\
        &= \sum_{i_c}\frac{A_{i_r i_c}}{\sqrt{d_{i_r i_c}}}\ket{0}_{a_e}\ket{i_c}_{a_c}\ket{i_r}_{\rm in} + \ket{\neq 0}_{a_e}\ket{\ldots}.
\end{split}
\end{equation}
After that, the oracle $O_M$ transfers the column indices from $a_c$ to the input register:
\begin{equation}\label{eq:OM}
\begin{split}
O_M&\left(\sum_{i_c}\frac{A_{i_r i_c}}{\sqrt{d_{i_r i_c}}}\ket{0}_{a_e}\ket{i_c}_{a_c}\ket{i_r}_{\rm in} 
    + \ket{\neq 0}_{a_e}\ket{\ldots}\right)\\
&= \sum_{i_c}\frac{A_{i_r i_c}}{\sqrt{d_{i_r i_c}}}\ket{0}_{a_e}\ket{i_r}_{a_c}\ket{i_c}_{\rm in} 
    + \ket{\neq 0}_{a_e}\ket{\ldots}.
\end{split}
\end{equation}
Finally, the oracles $O^\dagger_F O^{\rm B}_{F,{\rm corr}}$ entangle the states encoding the column indices in the input register with the zero state in the ancilla register:
\begin{equation}\label{eq:OFB}
\begin{split}
O^\dagger_F &O^{\rm B}_{F,{\rm corr}}\left(
    \sum_{i_c}\frac{A_{i_r i_c}}{\sqrt{d_{i_r i_c}}}\ket{0}_{a_e}\ket{i_r}_{a_c}\ket{i_c}_{\rm in} 
    + \ket{\neq 0}_{a_e}\ket{\ldots}
    \right)\\ 
&= \sum_{i_c}A_{i_r i_c}\ket{0}_{a_e}\ket{0}_{a_c}\ket{i_c}_{\rm in} 
    + \ket{\neq 0}_{a_e}\ket{\neq 0}_{a_c}\ket{\ldots}.
\end{split}
\end{equation}

% --------------------------------------------------------------------------------------
% --- STMC gates---
% --------------------------------------------------------------------------------------
\subsection{Single-target multicontrolled gates}\label{sec:STMC}
If one has a single-target gate $G$, whose matrix representation is
\begin{equation}
    \yG = 
    \begin{pmatrix}
        G_{00} & G_{01} \\
        G_{10} & G_{11}
    \end{pmatrix},
\end{equation}
then the corresponding STMC gate $C_{\{q_{c\delta}\}}G^{(q_t)}$ is defined as the gate $G$ acting on the target qubit $q_t$ and controlled by a set of qubits $\{q_{c\delta}\}$.
If $q_{c\delta}$ is a control qubit, then the gate $G$ is triggered if and only if $\ket{\delta}_{q_{c\delta}}$, where $\delta = 0$ or $1$.
If the gate $G$ acting on a quantum statevector $\ypsi$ of $n$ qubits is controlled by the qubit $q_{c\delta} \in [0,n)$, then only the statevector's elements with the indices $\{i_e\}_{q_{c\delta}}$ can be modified by the gate $G$:
\begin{subequations}\label{eq:control-elements}
\begin{eqnarray}
    &&i_e =  2 N_{\rm ctrl} j_b + j_{\rm step} + \delta N_{\rm ctrl},\\
    &&N_{\rm ctrl} = 2^{q_{c\delta}},\\
    &&j_b = [0,-1+2^n/(2N_{\rm ctrl})],\\
    &&j_{\rm step} = [0,N_{\rm ctrl}).
\end{eqnarray}
\end{subequations}
For instance, for $q_{c1} = 0$, every second element of $\ypsi$ can be modified by the gate $G$.
Then, the STMC gate $C_{\{q_{c\delta}\}}G^{(q_t)}$ can modify only the elements from the following set:
\begin{equation}\label{eq:set-control}
    \mathcal{S} = \bigcap_{q_{c\delta,k}\in \{q_{c\delta}\}} \{i_e\}_{q_{c\delta,k}},
\end{equation}
where $\bigcap$ is the intersection operator.
The most common case is when $q_t$ is a more significant qubit than the control ones, and the initial state of $q_t$ is the zero state. In this case, the action of $C_{\{q_{c\delta}\}}G^{(q_t)}$ can be described as
\begin{equation}
    \begin{split}
        &C_{\{q_{c\delta}\}}G^{(q_t)}\ket{0}_{q_t}
            \sum_{k\in\{i_c\}_{\rm init}}   \eta_{k}\ket{k}_{\rm ctrl}\\
            & = (G_{00}\ket{0}_{q_t} + G_{10}\ket{1}_{q_t}) 
                \sum_{k\in \{i_c\}_{\rm init}\cap \mathcal{S}}   \eta_{k}\ket{k}_{\rm ctrl}\\
            & + \ket{0}_{q_t}\sum_{k\in \{i_c\}_{\rm init}\setminus\mathcal{S}}  
                    \eta_{k}\ket{k}_{\rm ctrl},
    \end{split}
\end{equation}
where $\eta_{k}$ are the complex amplitudes of the initial state of the control register.

% --------------------------------------------------------------------------------------
% --- General algorithm ---
% --------------------------------------------------------------------------------------
\subsection{General algorithm for the block encoding}\label{sec:BE-general-idea}
To construct the BE oracle $U_A$, we use the following general procedure:
\begin{itemize}
    \item Normalize the matrix $\yA$ according to Eq.~\eqref{eq:first-normalization-A} using the the nonsparsity-related parameter~\eqref{eq:nonsparsity}.
    \item Introduce ancilla qubits necessary for intermediate computations in the oracle $U_A$ (Sec.~\ref{sec:ancillae}).
    \item Assume that the bitstring of the qubits of the input register (also called the state register) encodes a row index of the matrix $\yA$.
    \item Using STMC gates, construct the oracle $U_D$ following the idea presented in Eq.~\eqref{eq:UD-action} to encode column indices as a superposition of bitstrings in the state register (Sec.~\ref{sec:UD}).
    \item Compute the matrix $\yD^A$ or derive it using several matrices $\yD^A$ constructed for matrices $\yA$ of small sizes (Sec.~\ref{sec:D-large}).
    \item Using STMC rotation gates, construct the oracle $O_H$ to perform the transformation~\eqref{eq:OH-mapping} (Sec.~\ref{sec:OH}).
\end{itemize}
Once the circuit for the oracle $U_A$ is constructed using STMC gates, one can transpile the circuit into a chosen universal set of elementary gates.
Standard decomposition methods require at least $\oO(n)$ of basic gates\ycite{Barenco95}, where $n$ is the number of controlling qubits in an STMC gate.
Yet, it was recently shown\ycite{Claudon23} that it is possible to decompose an arbitrary STMC gate into a circuit with $\oO(\log_2(n)^{\log_2(12)}\log_2(1/\epsilon_{\rm STMC}))$ depth where $\epsilon_{\rm STMC}$ is the allowed absolute error in the approximation of the STMC gate.
In our assessment of the BE oracle's scaling below, we assume that the corresponding circuit comprises STMC gates not decomposed into elementary gates.

% --------------------------------------------------------------------------------------
% --- Ancilla qubits ---
% --------------------------------------------------------------------------------------
\subsection{Ancilla qubits for the block encoding}\label{sec:ancillae}

% ---------------------------------------------------------
\begin{figure}[!t]
\centering
\includegraphics[]{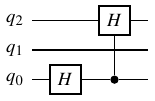}
\caption{\label{circ:Shpm1} 
    The circuit encoding the superposition of states $\ket{000}_q$, $\ket{001}_q$, and $\ket{101}_q$ for a given input zero state. 
}\end{figure}

\begin{figure}[!t]
\centering
\includegraphics[]{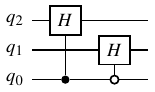}
\caption{\label{circ:Shp2} 
    The circuit encoding the superposition of states $\ket{000}_q$, $\ket{001}_q$, and $\ket{010}_q$ for a given input state produced by the circuit shown in Fig.~\ref{circ:Shpm1}. 
    To encode the superposition of states $\ket{100}_q$, $\ket{101}_q$, and $\ket{110}_q$, one uses the same circuit except that an additional $x$ gate is applied to the qubit $q_2$ in the end.
}\end{figure}

\begin{figure}[!t]
\centering
\includegraphics[]{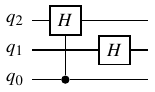}
\caption{\label{circ:Shp3} 
    The circuit encoding the superposition of states 
    $\ket{000}_q$, $\ket{001}_q$, $\ket{010}_q$, and $\ket{011}$ for a given input state produced by the circuit shown in Fig.~\ref{circ:Shpm1}.
    To encode the superposition of states $\ket{100}_q$, $\ket{101}_q$, $\ket{110}_q$, and $\ket{111}$, one uses the same circuit except that an additional $x$ gate is applied to the qubit $q_2$ in the end.
}\end{figure}
% ---------------------------------------------------------

The first step to construct a circuit of the BE oracle is to introduce ancilla qubits and assign the meaning to their bitstrings. 
As seen from Fig.~\ref{fig:A-colored-structure}, the matrix $\yA$ is divided into four submatrices.
To address each submatrix, we introduce the ancilla qubit $a_f$.
In combination with the register $r_f$ introduced in Eq.~\eqref{eq:psi-ket}, the register $a_f$ allows addressing each submatrix in the following way:
\begin{subequations}\label{eq:af}
\begin{eqnarray}
    &&\ket{0}_{a_f}\ket{0}_{r_f}\to \yF,\\
    &&\ket{1}_{a_f}\ket{0}_{r_f}\to \yC^E,\\
    &&\ket{0}_{a_f}\ket{1}_{r_f}\to \yC^f,\\
    &&\ket{1}_{a_f}\ket{1}_{r_f}\to \yS.
\end{eqnarray}
\end{subequations}
If $i_r$ and $i_c$ are row and column indices of $\yA$, respectively, such that
\begin{subequations}\label{eq:split-indices}
\begin{eqnarray}
    &&i_r = i_{fr} N_{xv} + i_{xr}N_v + i_{vr},\\
    &&i_c = i_{fc} N_{xv} + i_{xc}N_v + i_{vc},
\end{eqnarray}
\end{subequations}
with integers $i_{fr}, i_{fc} = [0,1]$, $i_{xr}, i_{xc} = [0, N_x)$, and $i_{vr}, i_{vc} = [0, N_v)$,
then the ancilla $a_f$ encodes the integer $i_{fc}$.

According to Fig.~\ref{fig:A-colored-structure}, each submatrix is split into $N_x^2$ blocks of size $N_v\times N_v$ each.
Within each submatrix, only the diagonal and a few off-diagonal (in the submatrix $\yF$) blocks contain nonzero matrix elements.
To address these blocks, we introduce the ancilla register $a_{xr}$ with three qubits.
This register stores the relative positions of blocks with respect to the main diagonal within each submatrix. 
The states $\ket{000}_{a_{xr}}$ and $\ket{100}_{a_{xr}}$ correspond to the blocks on the main diagonal of a submatrix.
The states $\ket{001}_{a_{xr}}$ and $\ket{010}_{a_{xr}}$ correspond to the off-diagonal blocks shifted by one and two blocks to the right, respectively.
The states $\ket{101}_{a_{xr}}$ and $\ket{110}_{a_{xr}}$ correspond to the off-diagonal blocks shifted by one and two blocks to the left, respectively.
Using same notation as in Eq.~\eqref{eq:split-indices}, the meaning of the register $a_{xr}$ can be described schematically as follows: 
\begin{subequations}\label{eq:axr}
\begin{eqnarray}
    &&\ket{000}_{a_{xr}}\ \text{and}\ \ket{100}_{a_{xr}} \to i_{xc} = i_{xr},\\
    &&\ket{001}_{a_{xr}} \to i_{xc} = i_{xr} + 1,\\
    &&\ket{010}_{a_{xr}} \to i_{xc} = i_{xr} + 2,\\
    &&\ket{101}_{a_{xr}} \to i_{xc} = i_{xr} - 1,\\
    &&\ket{110}_{a_{xr}} \to i_{xc} = i_{xr} - 2.
\end{eqnarray}
\end{subequations}
To encode the above bitstrings, one can use the circuits shown in Figs.~\ref{circ:Shpm1} and~\ref{circ:Shp2}.

In the submatrix $\yC^f$, some of the blocks have rows that contain $N_v$ nonzero elements.
To address these elements, we introduce the ancilla register $a_v$ with $n_v$ qubits. 
This register stores the integer $i_{vc}$ [Eq.~\eqref{eq:split-indices}] when the elements of $\yC^f$ and $\yC^E$ are addressed; otherwise, the register is not used.
For instance, to address the nonzero elements in the submatrix $\yC^f$, one uses the following encoding:
\begin{equation}\label{eq:av}
    \ket{0}_{a_f}\ket{i_{vc}}_{a_v}\ket{1}_{r_f}\ket{i_x}_{r_x}\ket{0}_{r_v} \to C^f_{i_x N_v,\ i_x N_v + i_{vc}}.
\end{equation}
As regards to $\yC^E$, since all its nonzero elements have $i_{vc} = 0$, one keeps the register $a_v$ in the zero state when $\ket{1}_{a_f}\ket{0}_{r_f}$.

The ancilla register $a_{vr}$ with $3$ qubits is introduced to encode positions of the matrix elements within the nonzero blocks of the submatrix $\yF$.
In particular, the states $\ket{000}_{a_{vr}}$ and $\ket{100}_{a_{vr}}$ correspond to the elements on the main diagonal of a block.
The states $\ket{001}_{a_{vr}}$, $\ket{010}_{a_{vr}}$, and $\ket{011}_{a_{vr}}$ correspond to the elements shifted by one, two and three cells, respectively, to the right from the diagonal.
The states $\ket{101}_{a_{vr}}$, $\ket{110}_{a_{vr}}$, and $\ket{111}_{a_{vr}}$ correspond to the elements shifted by one, two and three cells, respectively, to the left from the diagonal.
Using the notations from Eq.~\eqref{eq:split-indices}, the meaning of the register $a_{vr}$ can be described as 
\begin{subequations}\label{eq:avr}
\begin{eqnarray}
    &&\ket{000}_{a_{vr}}\ \text{and}\ \ket{100}_{a_{vr}} \to i_{vc} = i_{vr},\\
    &&\ket{001}_{a_{vr}} \to i_{vc} = i_{vr} + 1,\\
    &&\ket{010}_{a_{vr}} \to i_{vc} = i_{vr} + 2,\\
    &&\ket{011}_{a_{vr}} \to i_{vc} = i_{vr} + 3,\\
    &&\ket{101}_{a_{vr}} \to i_{vc} = i_{vr} - 1,\\
    &&\ket{110}_{a_{vr}} \to i_{vc} = i_{vr} - 2,\\
    &&\ket{111}_{a_{vr}} \to i_{vc} = i_{vr} - 3.
\end{eqnarray}
\end{subequations}
To encode the above bitstrings, one can use the circuits shown in Figs.~\ref{circ:Shpm1} and~\ref{circ:Shp3}.

Also, we introduce the ancilla $a_e$ whose zero-state's amplitude will encode the complex value of a given matrix element.

% ************************************************************************
% *** UD ***
% ************************************************************************
\subsection{Constructing the oracle $\mathbf{U_D}$}\label{sec:UD}

% -------------------------------------------------------------------------------------
% -------------------------------------------------------------------------------------
\begin{figure*}[!t]
\centering
\includegraphics[]{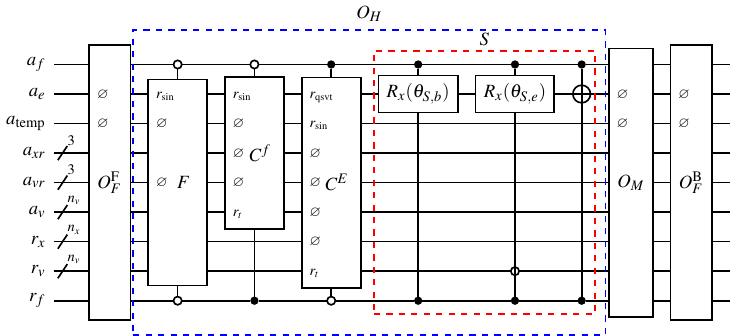}
\caption{\label{circ:BE-general} 
    The circuit representation of the BE oracle $U_A$.
    Here, $O^{\rm F}_F = O^{\rm F}_{F,{\rm corr}} O_F$ and $O^{\rm B}_F = O^\dagger_F O^{\rm B}_{F,{\rm corr}}$ are introduced using Eq.~\eqref{eq:UA-general}.
    The oracle $O_M$ is shown in Fig.~\ref{circ:OM}.
    The oracles $F$, $C^f$, $C^E$, and $S$ encode the corresponding submatrices introduced in Eq.~\eqref{eq:A}, and are described in Sec.~\ref{sec:OH}.
    Here, the symbol $\varnothing$ means that the gate does not use the corresponding qubit.
    The qubit $r_{\rm sin}$ shown in some oracles indicates that the corresponding oracle includes the circuit shown in Fig.~\ref{circ:sin}.
    The qubit $r_{\rm qsvt}$ indicates that the oracle $C^E$ is constructed using QSVT.
    The positions of the qubits $r_f$ and $a_f$ are changed for the sake for clarity.
}\end{figure*}
% -------------------------------------------------------------------------------------
% -------------------------------------------------------------------------------------

% -------------------------------------------------------------------------------------
% -------------------------------------------------------------------------------------
\begin{figure}[!t]
\centering
\includegraphics[]{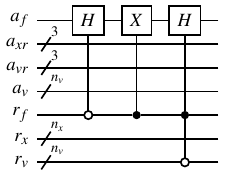}
\caption{\label{circ:OF-af} 
    The circuit to encode information into the ancilla $a_f$ according to Eq.~\eqref{eq:af}.
    The registers $r_v$, $r_x$, and $r_f$ encode the row index $i_r$, i.e. the indices $i_{vr}$, $i_{xr}$, and $i_{fr}$, correspondingly, according to Eq.~\eqref{eq:split-indices}.
}\end{figure}
% -------------------------------------------------------------------------------------
% -------------------------------------------------------------------------------------

% -------------------------------------------------------------------------------------
% -------------------------------------------------------------------------------------
\begin{figure}[!t]
\centering
\includegraphics[width=0.48\textwidth]{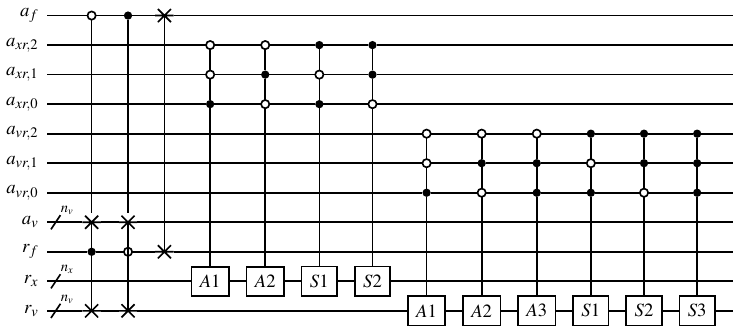}
\caption{\label{circ:OM} 
    The circuit of the oracle $O_M$.
    The adders $A1$-$A3$ and the subtractors $S1$-$S3$ are described in Appendix~\ref{app:sup-gates}.
}\end{figure}
% -------------------------------------------------------------------------------------
% -------------------------------------------------------------------------------------

A schematic of the BE oracle $U_A$ is shown in Fig.~\ref{circ:BE-general}, where the dashed blue box is the oracle $O_H$ described in Sec.~\ref{sec:OH}, and the rest of the boxes are components of the oracle $U_D$.

The oracle $U_D$ is built ad hoc manually by considering its suboracles separately.
The suboracle $O_F^{\rm F} = O^{\rm F}_{F,{\rm corr}} O_F$ is constructed based on Eq.~\eqref{eq:OFF}, where the `in' register includes the registers $r_v$, $r_x$, and $r_f$ described in Sec.~\eqref{sec:encoding-psi}.
The ancilla registers $a_v$, $a_{vr}$, $a_{xr}$, and $a_f$ described in Sec.~\ref{sec:ancillae} represent the $a_c$ register in Eq.~\eqref{eq:OFF}.
For instance, the part of the circuit $O_F$ encoding the ancilla $a_f$ [Eq.~\eqref{eq:af}] is shown in Fig.~\ref{circ:OF-af}, where the first Hadamard gate creates the address to the submatrices $\yF$ and $\yC^E$, and the Pauli $X$-gate generates the address to the elements of the submatrix $\yS$ with $i_{vr} > 0$ [using the notation from Eq.~\eqref{eq:split-indices}].
The last Hadamard gate produces the addresses to $\yC_f$ and $\yS$ at $i_{vr} = 0$.

The implementation of the suboracle $O_F^{\rm B} = O^\dagger_F O^{\rm B}_{F,{\rm corr}}$ is done based on Eq.~\eqref{eq:OFB}.
Its correcting part $O^{\rm B}_{F,{\rm corr}}$ is implemented ad hoc by varying positions and control nodes of STMC gates.

The suboracle $O_M$ which performs the mapping~\eqref{eq:OM} uses the SWAP gates to exchange the states in the registers $a_f$ and $a_v$ with the states in the registers $r_f$ and $r_v$, respectively, as shown in Fig.~\ref{circ:OM}.
To encode absolute column indices into the input registers using the states of the registers $a_{xr}$ and $a_{vr}$, one should follow the rules~\eqref{eq:axr} and~\eqref{eq:avr} and apply the quantum adders and subtractors described in Appendix~\ref{app:sup-gates}.
The important feature of the oracle $O_M$ is that the number of arithmetic operators in it does not depend on $N_x$ or $N_v$.
Since the circuits of the arithmetic operators scale as $\oO(n_x)$ or $\oO(n_v)$ depending on the target register of the operator, the scaling of $O_M$ is $\oO(n_x + n_v)$.
The full circuit of $U_D$ can be found in Ref.~\yocite{EVMcircuit}.
The modeling of the circuit is done using the QuCF framework\ycite{QuCF}, which is based on the QuEST toolkit\ycite{Jones19}.

\section{Construction of the oracle $\mathbf{O_H}$}\label{sec:OH}

% -------------------------------------------------------------------------------------------------------
% -------------------------------------------------------------------------------------------------------
\begin{figure*}[!t]
\centering
\includegraphics[width=0.90\textwidth]{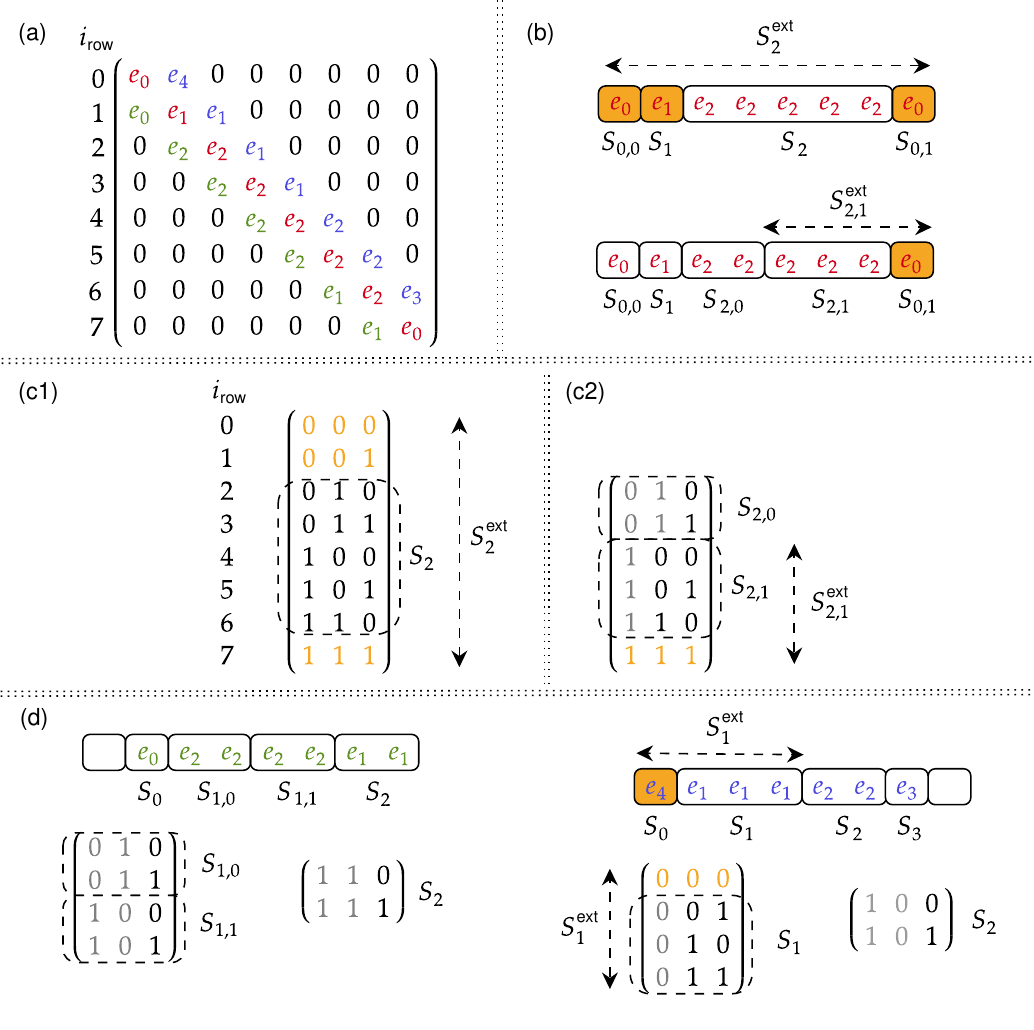}
\caption{\label{fig:schematic-OH} 
    A schematic showing decomposition of a matrix into sets for the construction of the oracle $O_H$. 
    (a): A simple matrix of size $N_A = 8$ is taken as an example. 
    This matrix consists of three diagonals marked in different colors, and each diagonal is considered separately. The element values are indicated as $e_i$, where $e_i \neq e_j$ if $i\neq j$.
    (b): A schematic showing how the main diagonal marked in red in (a) can be split in several sets where each set groups several matrix elements of the same value. 
    (c1): A matrix with bitstrings of the row indices of the original set $\yset_2$ and of its extended version $\ysete_{2}$.
    This extension results in three overlapped elements marked in orange, i.e. $\ysete_2$ overlaps elements of the sets $S_{0,0}$, $S_1$, and $S_{0,1}$.
    (c2): A bitstring matrix of the set $\yset_2$ which has been split into two sets, $\yset_{2,0}$ and $\yset_{2,1}$, where only the latter is extended. This extension results in a single overlapped element.
    (d): A schematic showing the splitting and extension (if necessary) of sets in the left and right diagonals.
    The empty cells indicate that the considered diagonal does not have elements at the corresponding rows.
    The bits indicated in gray in the bitstring matrices are chosen as the control nodes of the STMC gates encoding the extended sets (Fig.~\ref{circ:OH-example}). 
}\end{figure*}
% -------------------------------------------------------------------------------------------------------
% -------------------------------------------------------------------------------------------------------

% -------------------------------------------------------------------------------------------------------
% -------------------------------------------------------------------------------------------------------
\begin{figure*}[!t]
\centering
\includegraphics[width=0.90\textwidth]{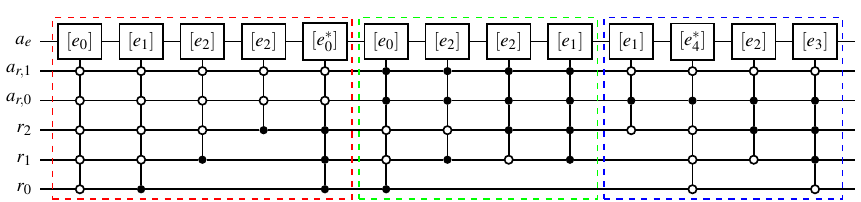}
\caption{\label{circ:OH-example} 
    The circuit of the oracle $O_H$ encoding the elements of the tridiagonal matrix from Fig.~\ref{fig:schematic-OH}.(a).
    The dashed blocks indicate the parts of the oracle encoding the main diagonal (the red box), the left diagonal (the green box), and the right diagonal (the blue box), correspondingly.
    Here, the register $r$ encodes the matrix row indices.
    The ancilla register $a_r$ has the similar meaning to that explained in Eq.~\ref{eq:axr} (although here the register has two qubits) and is used to address the matrix diagonals.
    The ancilla $a_e$ is initialized in the zero state, and the matrix elements' values are written into the amplitude of $\ket{0}_{a_e}$.
    The gate $[e_i]$ is a schematic representation of a rotation gate whose rotation angle is chosen to encode the value $e_i$ as explained in Eq.~\eqref{eq:OH-action}.
    The gate $[e_i^*]$ is a rotation gate whose angle is computed taking into account the overlapping of the element $e_i$ with one or several extended sets. 
    For instance, the third gate in the red box encodes the set $\yset_{2,0}$.
    The fourth gate encodes the extended set $\yset_{2,1}^{\rm ext}$. 
    Since $\ysete_{2,1}$ intersects with $S_{0,1}$, the angle for the fifth gate is computed taking into account the action of the fourth gate, as explained in Sec.~\ref{sec:OH-correction}.
}\end{figure*}
% -------------------------------------------------------------------------------------------------------
% -------------------------------------------------------------------------------------------------------

% ************************************************************************
% *** General strategy ***
% ************************************************************************
\subsection{General strategy}\label{sec:OH-strategy}
The purpose of the oracle $O_H$ is to compute $A_{i_r i_c}$ following Eq.~\eqref{eq:OH-action}. 
Therefore, further in the text, we consider the following rescaled matrix elements:
\begin{equation}\label{eq:rescaling-A}
    A_{i_r i_c} \to \frac{A_{i_r i_c}}{|d_{i_r i_c}|}.
\end{equation}

One can see from Sec.~\ref{sec:matrix} that two types of elements can be distinguished in $\yA$.
The first type includes elements that depend only on the discretization and system parameters such as the antenna frequency $\omega_0$ and the sizes of the grid cells.
These elements sit in the submatrix $\yS$ [Eq.~\eqref{eq:S}] and the submatrix $\yFF$ introduced after Eq.~\eqref{eq:FB}.
Since these elements appear mainly from the spatial, velocity, and time derivatives, they remain mostly constant at bulk spatial and bulk velocity points, but become highly intermittent at spatial and velocity boundaries.
% For instance, before the rescaling~\eqref{eq:rescaling-A}, all diagonal elements of $\yS$ have the same value.
% After the rescaling, all diagonal elements at $i_{vr} > 0$ keep the same value, but the elements at $i_{vr} = 0$ are modified. 
% The block encoding of $\yS$ is discussed in detail in Sec.~\ref{sec:S}.
The values of the elements of $\yFF$ strongly vary at boundaries of each $N_v\times N_v$ block.
These elements also change from one block to another depending on the original values of $\yFF$ [Eqs.~\eqref{eq:FL},~\eqref{eq:FR}, and~\eqref{eq:FB}] and also on the specific implementation of $U_D$, which influences the rescaling~\eqref{eq:rescaling-A}.
% Therefore, one needs an algorithm to analyze the matrix $\yFF$ and, based on this analysis, construct the corresponding quantum circuit.
% Such an algorithm is described in next sections. 

Another type of elements in $\yA$ are those that depend on velocity $v$ [and the background distribution function $H(v)$].
They form continuous profiles.
However, this continuity can be lost after the rescaling~\eqref{eq:rescaling-A}.
Such elements enter the submatrices $\yC^f$, $\yFP$, and $\yC^E$, whose block encoding is described in sections~\ref{sec:Cf},~\ref{sec:FP}, and~\ref{sec:CE}, respectively.

% ************************************************************************
% *** Block encoding the matrix FF ***
% ************************************************************************
\subsection{The algorithm for block encoding  $\yFF$}\label{sec:FF-oracle}

% *** Main idea ************************************************************************
\subsubsection{Main idea}
The main idea behind the algorithm for BE $\yFF$ is to decompose this matrix into sets of elements such that all elements would have the same value within each set.
After that, one extends large enough sets (as will be described further) in such a way that each extended set is encoded by a single STMC gate. 
However, after the extension, some sets may end up overlapping each other.
In this case, the elements in the intersections (i.e. in the overlapped regions) should be corrected by additional STMC gates.
The extension of some small sets that include a few matrix elements often leads to a significant overlapping with other sets. 
Therefore, it is often more efficient to encode a small set by encoding each element in it by its own STMC gate.
Also, the representation of a matrix by sets is not unique.
Because of that, we seek a minimal number of sets maximizing the number of large sets whilst minimizing the number of intersections and the number of small sets.
Ideally, such optimization should be done using, for instance, deep reinforcement learning\ycite{Sutton18}.
However, in the current version of the algorithm, the parameters are defined ad hoc.
% A schematic explaining the algorithm is depicted in Fig.~\ref{fig:schematic-OH}.

This algorithm can be performed in $N_{\rm iter}$ iterations:
\begin{equation}
    O_H \approx O_H^{(N_{\rm iter})}\ldots O_H^{(2)} O_H^{(1)},
\end{equation}
where $O_H^{(1)}$ is the initial version of the oracle $O_H$, $O_H^{(k)}$ is the correction to $O^{(1)}_H$ provided by the $(k>1)$-th iteration, and the final correction of overlapped elements and encoding of small sets is performed at the $N_{\rm iter}$-th iteration.
After the $k$-th iteration, one has:
\begin{equation}
    \delta \yA^{(k)} = \yA - \yA^{(k)},
\end{equation}
where the matrix $\yA^{(k)}$ is encoded by the sequence $O_H^{(k)}\ldots O_H^{(2)} O_H^{(1)}$.
If $\delta A^{(k)}_{i_r, i_c} = 0$, then the first $k$ iterations correctly encode the matrix element at the row $i_r$ and the column $i_c$.
If $|\delta A^{(k)}_{i_r, i_c}| > 0$, then the corresponding element should be corrected.
To do that, one considers the following matrix:
\begin{equation}
    \tilde{A}^{(k+1)}_{i_r, i_c} = 
    \left\{ \begin{aligned}
        A^{(k)}_{i_r, i_c}&,\quad |\delta A^{(k)}_{i_r, i_c}| > 0,\\
        0&,\quad \delta A^{(k)}_{i_r, i_c} = 0
    \end{aligned}\right.
\end{equation}
that provides information about which matrix elements still should be corrected.
At the $(k+1)$-th iteration, one decomposes $\tilde{\yA}^{(k+1)}$ into sets whose extension is restricted by the condition that the extended sets would not overlap the zero elements of $\tilde{\yA}^{(k+1)}$.
The $(k+1)$-th iteration constructs the oracle $O_H^{(k+1)}$ correcting $\yA^{(k)}$ by using $\tilde{\yA}^{(k+1)}$.
The correcting procedure is discussed in Sec.~\ref{sec:OH-correction}.
Because each extension of sets may lead to additional overlapped elements, it is not guaranteed that the $(k+1)$ iteration corrects all overlapped elements.
For that reason, at the last iteration, the matrix $\tilde{\yA}^{(N_{\rm iter})}$ is encoded element-by-element without extending the sets.
In other words, the $N_{\rm iter}$-th iteration corrects all remaining badly encoded matrix elements without introducing its own errors, which otherwise appear during the sets extension..
Finally, by varying the number $N_{\rm iter}$, one can construct the oracle $O_H$ with a near-optimal scaling with respect to the matrix size.

% *** Decomposition ************************************************************************
\subsubsection{Decomposition of the matrix}\label{sec:OH-matrix-decomposition}
The elements in the submatrix $\yFF$ are arranged along several diagonals 
(Fig.~\ref{fig:A-colored-structure}).
The number of these diagonals is less or equal to the nonsparsity of the matrix $\yA$ and depends neither on $N_x$ nor on $N_v$.
Therefore, the first step is to represent $\yFF$ as a group $\{\ydiag\}$ of separate diagonals $\ydiag$ [Fig.~\ref{fig:schematic-OH}.(a)] and then consider each diagonal $\ydiag$ separately.
One can specify a diagonal $\ydiag$ by using a constant integer $\Delta i$ as
\begin{equation}\label{eq:diagonal}
    \ydiag = \{\ysFF_{i_r, i_r + \Delta i} \big|\  
        i_r = [\delta_{{\rm sign}_{\Delta i}, -1}|\Delta i|,\ N_{xv} - \delta_{{\rm sign}_{\Delta i}, 1}\Delta i)  \}, 
\end{equation}
where ${\rm sign}_{\Delta i} = \Delta i / |\Delta i|$.
Each $\ydiag$ is encoded independently of other diagonals (as shown by various dashed blocks in Fig.~\ref{circ:OH-example}) that results in a linear dependence of the depth of the oracle $O_H$ on the nonsparsity $\varsigma_{\yFF}$ of the submatrix $\yFF$.

Each diagonal is represented by a group of sets $\{\yset\}$ where each set $\yset$ contains elements of the same value $\yvset$:
\begin{equation}
    \yset = \{\yvset, \{i_r\}|\ \yFF_{k, k + \Delta i} = \yvset, \forall k \in \{i_r\} \},
\end{equation}
where $\{i_r\}$ are the row indices of all matrix elements described by the set $\yset$.
Each set stores only $\yvset$ and $\{i_r\}$.
A set can be divided into smaller sets (Sec.~\ref{sec:OH-splitting}) and, thus, two different $\yset$ can have the same $\yvset$.

Once the matrix $\yFF$ is split into $\{\ydiag\}$, where $\ydiag = \{S\}$, one constructs a circuit representation of $O_H$.
The element-by-element encoding is inefficient, because it will require $\oO(N_{xv})$ quantum gates.
The main purpose of our algorithm is to construct a circuit with the scaling better than $\oO(N_{xv})$.
Ideally, the scaling should be $\oO(\text{polylog}(N_{xv})\varsigma_{\yFF})$.
[As shown in Ref.~\yocite{Novikau23}, the QSVT for solving a multi-dimensional stationary wave problem described by a matrix of size $N$ and nonsparsity $\varsigma$ can have a polynomial speedup in comparison to classical conjugate-gradient-based algorithms if the corresponding block-encoding oracle scales as $\oO(\text{polylog}(N)\varsigma)$].

To achieve a better scaling, the algorithm encodes the sets instead of the matrix elements by using the fact that each $\yset$ contains elements with the same value.
The simplest way to encode $\yset$ is to use a quantum arithmetic comparator\ycite{Suau21, Novikau23} to find whether $i_r \in \{i_r\}_{\yset}$, where $i_r$ is the row index encoded into the registers $r_v$, $r_x$, and $r_f$ (Sec.~\ref{sec:encoding-psi}).
However, for each $\yset$, one will need to use at least two comparators, and each comparator requires two ancilla qubits (although the ancillae can be potentially reused by various comparators).
Instead, we try to arrange sets in such a manner that each large enough set would be encoded by a single STMC gate.
Such an arrangement is performed by extending $\yset$.

% *** Extenstion ************************************************************************
\subsubsection{Sets extension}
To encode all elements of $\yset$ at once, the algorithm extends $\yset$ in such a way that the extended set $\ysete$ is computed by a single STMC gate.
To make it possible, it is necessary to complete $\{i_r\}$ of $\yset$ by another set $\{i_r^{\rm com}\}$ such that the resulting set $\{i_r\}\cup\{i_r^{\rm com}\}$ could be represented by Eqs.~\eqref{eq:control-elements} and ~\eqref{eq:set-control}.
In other words, $\yset$ is extended to ensure that it would be possible to find a STMC gate $C_{\{q_{c\delta}\}}G^{(a_e)}$ such that $G_{00} = \yvset$ and $\mathcal{S} \equiv \{i_r\}\cup\{i_r^{\rm com}\}$.

The extended set $\ysete$ consists of a core, which is the original nonextended set $\yset$, and a complement $\ysetc$: 
$\ysete = \yset \cup \ysetc$ and $\yset \cap \ysetc = \varnothing$, where $\ysetc$ includes $N_c$ elements described by the row indices $\{i_r^{\rm com}\}$:
\begin{equation}
    \ysete = \{\yvset, \{i_r\}\cup\{i_r^{\rm com}\}|\ \ysFF_{k, k + \Delta i} = \yvset, \forall k \in \{i_r\} \}.
\end{equation}
Note here that the matrix elements $\ysFF_{j, j + \Delta i}$ for $j \in \{i_r^{\rm com}\}$ may have a value different from $\yvset$ but the extended set $\ysete$ assumes that they do have the value $\yvset$.
Thus, the extended sets can assign wrong values to some matrix elements.
In this case, we speak about overlapping of several sets.
The cores of extended sets never intersect, $S_j \cap S_k = \varnothing$ for $j\neq k$, but the complements can overlap with the cores or complements of other sets. 
Hence, one can have only complement--complement and complement--core overlapping.
The matrix elements that sit in the overlapped regions of several sets should be corrected.
The correction employs supplemental STMC gates. 
Therefore, it is important to minimize the number of overlapped elements.
The correction can be done in different ways, and a possible algorithm for that is described in Sec.~\ref{sec:OH-correction}.

We should also note that, sometimes, small sets are extended to significantly larger sets, thus, drastically increasing the number of overlapped elements.
Because of that, it is better to encode small enough sets such that each element in them is computed by a separate STMC gate.

In well-structured matrices, such as those that appear in classical linear wave problems, the number of sets is significantly less than the number of matrix elements.
Since each extended set is encoded by a single STMC gate, the scaling of the resulting oracle should be significantly better than in the case when the matrix is encoded element by element.
For instance, let us assume that a diagonal $\ydiag$ consists of $N$ elements.
If all these elements are equal to each over, then $\ydiag$ comprises a single set, and this set can be encoded into a quantum circuit by using a single STMC rotation gate.
Another case is when there are $N_S$ elements of the same value where $(N - N_S) \ll N$, then $\ydiag$ has a dominant set with $N_S$ elements.
This set can be extended to the whole diagonal as shown in Fig.~\ref{fig:schematic-OH}.(b) for the set $\yset_2$.
In this case, the complement $\ysetc$ contains $(N-N_S) \ll N$ elements and each of these elements are encoded by a separate STMC gate.
This means that the oracle $O_H$ encoding the diagonal $\ydiag$ has $(N-N_S + 1) \ll N$ STMC gates.

However, extending the original set may not always be the best approach. 
Instead, $\yset$ can be split (Sec.~\ref{sec:OH-splitting}) into several subsets and then each subset can be extended individually. 
As demonstrated in Fig.~\ref{fig:schematic-OH}.(b) for the set $\yset_2$, the splitting allows to reduce the number of overlapped elements from $N_c = 3$ to $N_c = 1$.

% ************************************************************************
% *** Correction ***
% ************************************************************************
\subsubsection{Correction of overlapped elements}\label{sec:OH-correction}
If two sets intersect with each other, the overlapped elements should be corrected by supplemental gates.
Let us consider the correction on the example from Fig.~\ref{fig:schematic-OH}, where, for instance,
two sets, $S_{2,1}^{\rm ext}$ and $S_{0,1}$, overlap at the element $e_0$.
According to Fig.~\ref{fig:schematic-OH}.(c2), to encode the set $S_{2,1}^{\rm ext}$, one needs a single STMC gate controlled by the most significant qubit of the input register $r$ (as shown in Fig.~\ref{circ:OH-example} by the fourth gate in the red box).
To encode an arbitrary complex value, one can use the gate $R_c(2\theta_{z,2}, 2\theta_{y,2})$ described in Appendix~\ref{app:sup-gates}, which should act at the zero state of the ancilla $a_e$.
In particular, to compute the value $e_2$, one needs to chose the gates' parameters in such a way that $e_2 = \cos(\theta_{y,2})e^{-\yi\theta_{z,2}}$.
In this case, one obtains:
\begin{equation}\label{eq:set-21}
    \ket{\psi_{\rm ext}}_{a_e} = R_c(2\theta_{z,2}, 2\theta_{y,2})\ket{0}_{a_e} = e_2 \ket{0}_{a_e} + w_2 \ket{1}_{a_e},
\end{equation}
where $w_2$ is a complex value defined by the angles of the gate $R_c$.
Thus, two components of the state $\psi_{\rm ext}$ are $\psi_{\rm ext,0} = e_2$ and $\psi_{\rm ext,1} = w_2$.
This gate controlled by the most significant qubit of the register $r$ (Fig.~\ref{circ:OH-example}) 
entangles the state $\ket{\psi_{\rm ext}}_{a_e}$ with the states $\ket{i_r}_{r}$ where $i_r = [4,7]$.
However, at $i_r = 7$, the main diagonal has an element with the value $e_0$ instead of $e_2$.
This means that a supplemental gate $R_c(2\theta_{z,c}, 2\theta_{y,c})$ should act on the state $\ket{\psi_{\rm ext}}_{a_e}\ket{7}_{r}$ to encode the value $e_0$:
\begin{equation}\label{eq:OH-correction}
    \ket{\psi_{\rm res}}_{a_e} = R_c(2\theta_{z,c}, 2\theta_{y,c}) \ket{\psi_{\rm ext}}_{a_e} = e_0 \ket{0}_{a_e} + w_0 \ket{1}_{a_e},
\end{equation}
where we are not interested in the value $w_0$.
This explains the last gate in the red box in Fig.~\ref{circ:OH-example}, denoted $[e_0^*]$.

The equation~\eqref{eq:OH-correction} leads to the following equations for $\theta_{z,c}$ and $\theta_{y,c}$:
\begin{subequations}\label{eq:correction}
\begin{eqnarray}
    &&g_{00}\cos\theta_{y,c}  - g_{01}\sin\theta_{y,c} = \psi_{\rm res,0}^{\rm real},\\
    &&g_{10}\cos\theta_{y,c}  - g_{11}\sin\theta_{y,c} = \psi_{\rm res,0}^{\rm imag},
\end{eqnarray}
\end{subequations}
where  
\begin{subequations}\label{eq:correction-add}
\begin{eqnarray}
    &&g_{00} = \psi_{\rm ext,0}^{\rm real}\cos\theta_{z,c} + \psi_{\rm ext,0}^{\rm imag}\sin\theta_{z,c},\\
    &&g_{01} = \psi_{\rm ext,1}^{\rm real}\cos\theta_{z,c} - \psi_{\rm ext,1}^{\rm imag}\sin\theta_{z,c},\\
    &&g_{10} = \psi_{\rm ext,0}^{\rm imag}\cos\theta_{z,c} - \psi_{\rm ext,0}^{\rm real}\sin\theta_{z,c},\\
    &&g_{11} = \psi_{\rm ext,1}^{\rm real}\sin\theta_{z,c} + \psi_{\rm ext,1}^{\rm imag}\cos\theta_{z,c}.
\end{eqnarray}
\end{subequations}
In a more general case, when $N_g$ gates act on the same element $e_{\rm res}$, then one can find $\theta_{y,c}$ and $\theta_{z,c}$ in Eqs.~\eqref{eq:correction}-\eqref{eq:correction-add} from
\begin{subequations}
\begin{eqnarray}
    &&\psi_{\rm ext} = R_{c,N_g-1} R_{c, N_g-2}\dots R_{c,1} R_{c,0}\ket{0}_{a_e},\\
    &&\psi_{\rm res, 0} = e_{\rm res}.
\end{eqnarray}
\end{subequations}

% *** Splitting ************************************************************************
\subsubsection{Splitting sets}\label{sec:OH-splitting}
As we have already mentioned earlier, to minimize the number of overlapped elements, it may be more efficient to split sets such that the divided sets would have complements of smaller sizes and, thus, would require a smaller number of operations to correct the overlapped elements.
Such splitting and reorganization of sets can be done in many different ways and, ideally, should be delegated to machine learning, which is likely to find a more optimal set organization than one can feasibly do manually ad hoc.
However, there is also a simpler way to rearrange the sets, which is as follows.
For each set, one constructs a matrix $\yA_{\rm bits}$ with bitstrings representing the row indices of the set's elements as illustrated in Fig.~\ref{fig:schematic-OH}.(c) and Fig.~\ref{fig:schematic-OH}.(d).
The size of $\yA_{\rm bits}$ is $N_S\times n_b$, where $N_S$ is the set's size (the number of elements in the set), and $n_b = \log_2(N_A)$, where $N_A$ is the size of the matrix to be described by $O_H$ 
(in our case, the matrix is $\yFF$ and $N_A = N_{xv}$).

The rows in $\yA_{\rm bits}$ are enumerated by the index $r_b \in [0,N_S)$ and store the bitstring representations of the rows of the matrix $\yFF$.
Each row $r_b$ has $n_b$ cells with bits of different significance, where the leftmost cell stores the most significant bit.
Each column $c_b \in [0, n_b)$ in $\yA_{\rm bits}$ contains an array of bits of equal significance [Fig.~\ref{fig:schematic-OH}.(c)].
The column $c_b = 0$ contains the most significant bits.

To decide how to split the set, one checks the leftmost $N_{\rm check}$ columns of $\yA_{\rm bits}$.
Let us assume that the leftmost column, $c_b = 0$, has the bit $\delta$ at $r_b = 0$.
If all bitcells at $c_b = 0$ contain the same bit $\delta$, then one checks the bits in the next column. 
If $N_{\rm check}$ columns have bitcells with unchanging bits, then the set remains undivided.

However, if the column $c_b < N_{\rm check}$ has a bit $\delta$ at $r_b = 0$, but this bit changes at $r_{b,1}$,
then one splits $\yA_{\rm bits}$ into two matrices, $\yA_{\rm bits, 0}$ and $\yA_{\rm bits, 1}$.
The former matrix is the part of $\yA_{\rm bits}$ with the rows $[0, r_{b,1})$, the latter matrix is the part of $\yA_{\rm bits}$ with the row $[r_{b,1}, N_S)$.
These new matrices of bitstrings represent new sets.
These sets still describe matrix elements with the same value, but now these elements are combined into two sets instead of the original single set.
In this manner, $S_2$ in the main diagonal in Fig.~\ref{fig:schematic-OH} is split into two sets, $S_{2,0}$ and $S_{2,1}$.

One can vary $N_{\rm check}$ to find an optimal number of sets in $\ydiag$, but usually it is better to keep this number significantly smaller than $n_b$ because, otherwise, one can end up with a large number of small sets.
This happens because the bits of low significance have a higher possibility to change within a set.

% ************************************************************************
% *** Extension to larger matrices ***
% ************************************************************************
\subsection{Extrapolation of the matrix $\yD^A$}\label{sec:D-large}

To construct the oracle $O_H$, one needs to compute the matrix $\yD^A$ related to the oracle $U_D$ [Eq.~\eqref{eq:UD}] to perform the rescaling~\eqref{eq:OH-action}.
The parts of $\yD^A$ related to the submatrices $\yC^E$, $\yC^f$ and $\yS$ have a trivial structure and can be predicted for any $N_x$ and $N_v$.
For instance, $D^A_{N_{xv} + i_x N_v + i_v,\ N_{xv}+ i_x N_v + i_v} = 1$ for $i_x = [0, N_x)$ and $i_v = [1, N_v)$.
These elements correspond to the main diagonal of the submatrix $\yS$, where the matrix nonsparsity is $1$.
In the rows with indices $N_{xv} + i_x N_v$, the column indices of the nonzero elements are computed by a single Hadamard gate acting on the ancilla $a_f$ and by $n_v$ Hadamard gates acting on all qubits of the register $a_v$.
[These Hadamards gates appear in both $O_F$ and $O_F^\dagger$, according to Eq.~\eqref{eq:UD} and Fig.~\ref{circ:OF-af}.]
Because of that, $D^A_{N_{xv} + i_x N_v,\ N_{xv}+ i_x N_v} = 1/2$ that corresponds to the diagonal elements at $i_v = 0$ in the submatrix $\yS$, and $D^A_{N_{xv} + i_x N_v,\ i_x N_v + i_v} = 1/2^{n_v/2+1}$ with $i_v = [0, N_v)$ that corresponds to the nonzero elements of the submatrix $\yC^f$. 

Let us denote the upper left $N_{xv}\times N_{xv}$ part of $\yD^A$ as $\yD^F$.
The submatrix $\yD^F$ contains elements $d_{i_r i_c}$ by which the submatrix $\yF$ should be rescaled in Eq.~\eqref{eq:OH-action}.
The computation of $\yD^F$ is numerically challenging for a large $N_{xv}$.
Instead, $\yD^F$ of a large size can be extrapolated by using known matrices $\yD^F$ of several small sizes.
To do that, one creates a template of $\yD^F$ by using computed $\yD^F$ for small $N_x$ and $N_v$.
This can be done because the relative positions of the submatrix elements and the elements' values do not change with the increase of $N_{xv}$.
The template is used to reconstruct $\yD^F$ for any $N_{xv}$.
The algorithm for the creation of this template is the following.

Let us consider the matrix $\yD^F$ that is split into several diagonals in the same way as described in Sec.~\ref{sec:OH-matrix-decomposition}.
The algorithm considers each of these diagonals independently and assumes that the row index $r$ of this matrix depends on $N_{\rm ind}$ indices $r_l$:
\begin{equation}\label{eq:matrix-row-index-general}
    r = \sum_{l=0}^{N_{\rm ind}-1} r_l \prod_{j=(l+1)}^{N_{\rm ind}} N_j,
\end{equation}
where each index $r_l = [0, N_l)$ corresponds to the $l$-th dimension of the size $N_l$, and $N_{N_{\rm ind}} = 1$.
For instance, any matrix describing dynamics in 3-D real space ($N_{\rm ind} = 3$) has $r = r_0 N_y N_x + r_1 N_x + r_2$.
In our case, $r_0 = i_{xr}$ and $r_1 = i_{vr}$ according to Eq.~\eqref{eq:split-indices}.
Also, we define the following ordered set of indices:
\begin{equation}
    \yoset{l} = \{r_j \big|\ r_j = [0, N_j),\ j = [0, l) \}.
\end{equation}
The ordered set $\yoset{l}$ is empty if $l = 0$.

Let us consider the diagonal $\mathcal{D}$ defined in Eq.~\eqref{eq:diagonal}, where instead of $\yFF$ we now take $\yD^F$.
The vector with all matrix elements from $\mathcal{D}$ is denoted $\yM$.
It has $N_{xv}$ elements, where we formally define $M_{r} = \varnothing$ if the diagonal does not have any element at the row $r$.
Due to the dependence~\eqref{eq:matrix-row-index-general}, $\yM$ has a nested structure with $N_0$ blocks, where each block has $\prod_{j=1}^{N_{\rm ind}} N_j$ rows.
These rows can be combined into $N_1$ sub-blocks, where each sub-block has $\prod_{j=2}^{N_{\rm ind}} N_j$ rows, and so on.
Let us call all blocks of an equal size a layer.
There are $N_{\rm ind}$ layers.
The $i_l$-th block in the $l$-th layer is defined as
\begin{equation}\label{eq:blocks-in-layer}
    B_{\yoset{l}, i_l} = \bigg\{B_{\yoset{l} \cup i_l, i_{l+1}}\bigg\},
\end{equation}
where $i_l = [0, N_l)$, $l = [0, N_{\rm ind})$, and each block of the $(N_{\rm ind} - 1)$th layer is a single matrix element:
\begin{equation}
    B_{\yoset{q}, i_q} = \bigg\{M_{r} \big|\  r = \sum_{r_l\in\yoset{q}} r_l \prod_{j=(l+1)}^{q} N_j + i_q \bigg\},
\end{equation}
where $q = N_{\rm ind} - 1$.
Hence, the $l$-th layer has $N_l$ blocks, and each of these blocks has $\prod_{j=l+1}^{N_{\rm ind}} N_j$ matrix elements.

Identical adjacent blocks in the $l$-th layer are combined into blocksets.
A blockset in the $l$-th layer is denoted $\mathcal{B}_l$ and is defined as
\begin{eqnarray}\label{eq:blockset}
    \mathcal{B}_l = \bigg\{B_{\yoset{l}, r_b}, r_b, r_e\big|\ 
        B_{\yoset{l}, r} = B_{\yoset{l}, r_b}, \forall r = [r_b, r_e) \bigg\},
\end{eqnarray}
% where each blockset in the last layer (i.e. in the layer with the index $q = N_{\rm ind} - 1$) 
where each blockset in the $q = (N_{\rm ind} - 1)$th layer combines identical neighbor matrix elements
\begin{equation}\label{eq:blockset-last-layer}
    \mathcal{B}_q = \bigg\{M_{r}, r_b, r_e\big|\ M_{r} = M_{r_b}, \forall r = [r_b, r_e)\bigg\}.
\end{equation}
Since each block in the $l$-th layer includes blocks from the $(l+1)$th layer [Eq.~\eqref{eq:blocks-in-layer}], each block can be described by a group of blocksets:
\begin{equation}\label{eq:block-as-blockset}
    B_{\yoset{l},i_l} = \{\mathcal{B}_{l+1}\}_{i_l},
\end{equation}
where we keep the index $i_l$ to indicate that $\{\mathcal{B}_{l+1}\}_{i_l}$ represents the $i_l$-th block at the $l$-th layer. 
If the block $B_{\yoset{-1},i_{-1}}$ is formally  defined as the whole diagonal $\yM$, then $\{\mathcal{B}_{0}\}_{i_{-1}}$ is a group of blocksets where the blocks of the zeroth layer are sorted.
By combining Eqs.~\eqref{eq:blockset} and~\eqref{eq:block-as-blockset}, one can see that each blockset in the $l$-th layer can be defined as a group of blocksets from the $(l+1)$th layer:
\begin{equation}\label{eq:blockset-as-blocksets}
    \mathcal{B}_l = \bigg\{ \{\mathcal{B}_{l+1}\}_{r_b}, r_b, r_e\big|\ 
        \{\mathcal{B}_{l+1}\}_{r} = \{\mathcal{B}_{l+1}\}_{r_b}, \forall r = [r_b, r_e) \bigg\}.
\end{equation}
Thus, the whole diagonal $\yM$ can be represented as nested blocksets.
Representing a matrix by a group of diagonals comprising blocksets can be regarded as creating a template that represents a compressed image of the matrix.

In the matrix $\yD^F$, the number of diagonals $\mathcal{D}$ does not change with $N_{xv}$.
The matrix elements $M_r$ stored in the blocksets~\eqref{eq:blockset-last-layer} are also independent of $N_{xv}$.
(Some matrix elements of $\yD^A$ do change with $N_v$ because the nonsparsity of $\yC^f$ and $\yC^E$ changes with $N_v$.
Yet, the nonsparsity of the submatrix $\yF$ do not change with $N_{xv}$, and because of that, the elements of $\yD^F$ do not change as well.)
The indices $r_b$ and $r_e$ in Eqs.~\eqref{eq:blockset-as-blocksets} and~\eqref{eq:blockset-last-layer} depend linearly on $N_l$:
\begin{subequations}\label{eqs:sets-boundaries}
\begin{eqnarray}
&&r_b = \alpha_b + \beta_b N_l,\\
&&r_e = \alpha_e + \beta_e N_l,
\end{eqnarray}
\end{subequations}
where $l=[0,N_{\rm ind})$.
The unknown coefficients $\alpha_b$, $\alpha_e$, $\beta_b$, and $\beta_e$ are different for different blocksets.
To compute these coefficients, one constructs the templates for $N_{\rm ind}+1$ matrices where each matrix has at least one $N_l$ different from the corresponding $N_l$ of all other matrices.
For each of these computed $(N_{\rm ind}+1)$ templates, one knows the row indices $r_b$ and $r_e$ for each blockset.
One substitutes these indices to Eqs.~\eqref{eqs:sets-boundaries} from where the unknown coefficients now can be calculated.

Once the coefficients are computed, one can construct $\yD^F$ of an arbitrary size $N_{xv}$.
The equations~\eqref{eqs:sets-boundaries} allow to compute the positions $r_b$ and $r_e$ for all blocksets in $\yD^F$ for the required $N_{xv}$.

For the matrix $\yD^F$, $N_{\rm ind}$ equals two, where $N_0 = N_x$ an $N_1 = N_v$.
Thus, one needs to precalculate three matrices $\yD^F$ with various $N_x$ and $N_v$ to use Eqs.~\eqref{eqs:sets-boundaries}.
Yet, one should keep in mind that $N_x$ and $N_v$ should be large enough to take into account the influence of the boundary elements in $\yD^F$ properly.
More precisely, one should consider only matrices with $n_x \geq 4$ and $n_v \geq 4$.

% -------------------------------------------------------------------------------------------------------
% -------------------------------------------------------------------------------------------------------
\begin{figure}[!t]
\centering
\includegraphics[width=0.49\textwidth]{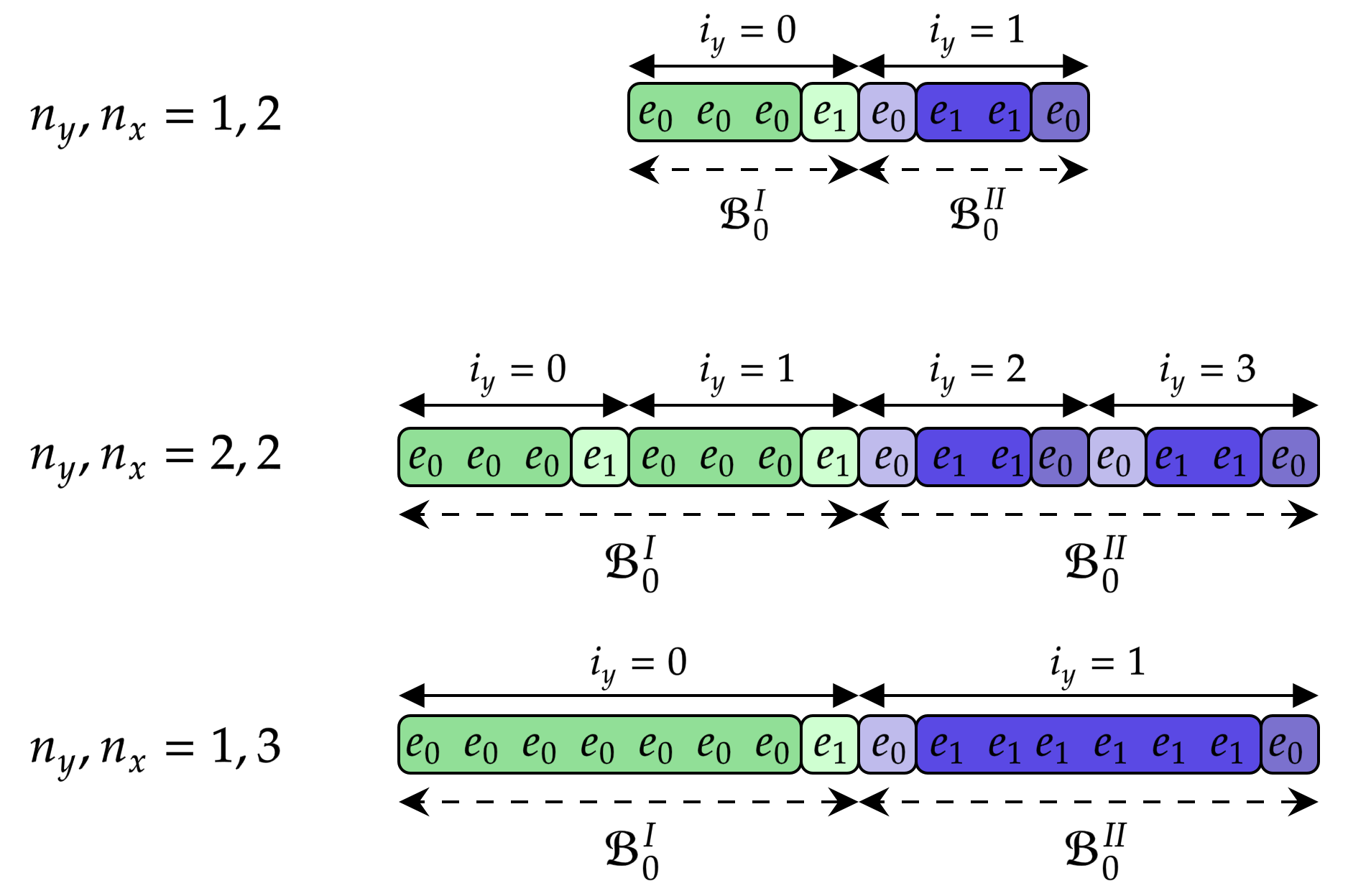}
\caption{\label{fig:schematic-matrix-extrapolation} 
    A schematic showing elements in the main diagonal of the matrix $\yD^{\rm ex}$ described in Eq.~\eqref{eq:example-D} for various $N_x = 2^{n_x}$ and $N_y = 2^{n_y}$.
    The matrix has $N_y$ blocks with $N_x$ elements each.
    These blocks are grouped into two blocksets, $\mathcal{B}_0^I$ and $\mathcal{B}_0^{II}$.
    The elements in each block of the blockset $\mathcal{B}_0^I$ are combined into two blocksets indicated by different shades of green.
    The elements in each block of the blockset $\mathcal{B}_0^{II}$ are combined into three blocksets indicated by different shades of blue. 
}\end{figure}
% -------------------------------------------------------------------------------------------------------
% -------------------------------------------------------------------------------------------------------

As an example illustrating the decomposition of a matrix into blocksets, let we consider a diagonal matrix $\yD^{\rm ex}$ with diagonal elements $D^{\rm ex}_{i, i}$, where $i = i_y N_x + i_x$ with $i_y = [0,N_y)$ and $i_x = [0,N_x)$ for $N_y = 2^{n_y}$ and $N_x = 2^{n_x}$:
\begin{subequations}\label{eq:example-D}
\begin{eqnarray}
&&D^{\rm ex}_{i, i} = e_0,\ i_y = [0,N_y/2),\ i_x = [0,N_x-2],\\
&&D^{\rm ex}_{i, i} = e_1,\ i_y = [0,N_y/2),\ i_x = N_x-1,\\
&&D^{\rm ex}_{i, i} = e_0,\ i_y = [N_y/2,N_y),\ i_x = 0\text{ and } N_x-1,\\
&&D^{\rm ex}_{i, i} = e_1,\ i_y = [N_y/2,N_y),\ i_x = [1,N_x-2].
\end{eqnarray}
\end{subequations}
Here, $e_0$ and $e_1$ are some nonequal values.
Due to the dependence on the indices $i_x$ and $i_y$, the main diagonal has a nested structure with $N_y$ blocks and with $N_x$ elements in each block. 
Other diagonals are empty (the matrix elements in these diagonals are equal to zero).
The main diagonal for different $n_y$ and $n_x$ is shown in Fig.~\ref{fig:schematic-matrix-extrapolation}.

In the zeroth layer, the diagonal is split into two blocksets, $\mathcal{B}_0^I$ and $\mathcal{B}_0^{II}$, independently of the matrix size.
In the next layer, $\mathcal{B}_0^I$ contains two blocksets (which are indicated in different shades of green in Fig.~\ref{fig:schematic-matrix-extrapolation}), and $\mathcal{B}_0^{II}$ contains three blocksets (indicated by different shadows of blue).
% The blockset $\mathcal{B}_0^{I}$ represents all $N_y/2$ identical blocks in the first half of the diagonal.
% The blockset $\mathcal{B}_0^{I}$ in its turn includes two blocksets: one describes the neighbor elements with the value $e_0$, another one includes the single element $e_1$.
Although the number of blocks and elements in $\yD^{\rm ex}$ changes with $N_x$ and $N_y$, the number of blocksets in any layer does not.
For instance, independently of $N_y$, the blockset $\mathcal{B}_0^{I}$ always contains only two blocksets.
Indeed, according to Eq.~\eqref{eq:blockset}, $\mathcal{B}_0^{I}$ stores a single copy of one of $N_y/2$ identical blocks in the first half of the diagonal. 
In turn, this single copy contains two blocksets, one of which saves a single copy of the elements $e_0$, and another saves the element $e_1$.
 
The indices stored by the blockset $\mathcal{B}_0^{I}$ are 
\begin{equation}
    r_b = 0,\quad r_e = \frac{1}{2} N_y.
\end{equation}
The indices of the first inner blockset of $\mathcal{B}_0^{I}$ are $r_b = 0$ and $r_e = N_x-2$.
The indices of the second inner blockset are $r_b = N_x-2$ and $r_e = N_x-1$.

The blockset $\mathcal{B}_0^{II}$ have the following boundaries:
\begin{equation}
    r_b = \frac{1}{2} N_y,\quad r_e = N_y.
\end{equation}
The indices of the inner blocksets of $\mathcal{B}_0^{II}$ can be easily found from Fig.~\ref{fig:schematic-matrix-extrapolation} or Eqs.~\eqref{eq:example-D}.

\subsection{Encoding the submatrix $\yS$}\label{sec:S}
To encode the submatrix $\yS$ described in Eq.~\eqref{eq:S}, we use the two gates $R_x$ with the following angles:
\begin{subequations}\label{eq:theta-S}
\begin{eqnarray}
    &&\theta_{S,b} = -2 \arcsin(\omega_0),\\
    &&\theta_{S,e} = -2 \arcsin(2 \omega_0) - \theta_{S,b}.\label{eq:theta-S-e}
\end{eqnarray}
\end{subequations}
% The factor $2$ under the arcsin in Eq.~\eqref{eq:theta-S-e} appears due to the rescaling~\eqref{eq:OH-action} as explained in Sec.~\ref{sec:D-large}.
As one can see from the red dashed box in Fig.~\ref{circ:BE-general}, the elements of $\yS$ are encoded by three gates, $R_x(\theta_{S,b})$, $R_x(\theta_{S,e})$, and $X$, applied to the ancilla $a_e$.
The first gate $R_x(\theta_{S,b})$ encodes $\yi\omega_0$ into the amplitude of the zero state of the ancilla $a_e$.
Due to the rescaling~\eqref{eq:rescaling-A}, the value $\yi\omega_0$ of each element at $i_{vr} = 0$ in $\yS$ is multiplied by the factor $2$.
This is taken into account by the second gate, $R_x(\theta_{S,e})$, which corrects the action of $R_x(\theta_{S,b})$.

% ************************************************************************
% *** Submatrix Cf ***
% ************************************************************************
\subsection{Encoding the submatrix $\yC^f$}\label{sec:Cf}

\begin{figure}[!t]
\centering
\includegraphics[]{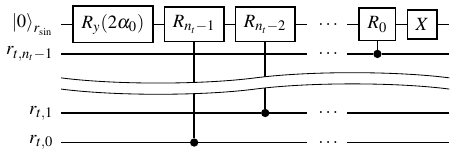}
\caption{\label{circ:sin} 
    The circuit encoding $\sin(\phi_i)$ according to Eq.~\eqref{eq:sin-gate}. 
    Here, $R_k = R_y(2\alpha_1/2^k)$.
}\end{figure}

To encode the matrix elements of $\yC^f$ [Eq.~\eqref{eq:Cf}] that depend on the velocity, we use the circuit shown in Fig.~\ref{circ:sin}.
It encodes the function $\sin(\phi_i)$ for $i = [0, N_t)$ with $N_t = 2^{n_t}$:
\begin{equation}\label{eq:sin-gate}
    \phi_i = \alpha_0 + i \Delta\phi,
\end{equation}
where $n_t$ is the number of qubits in the register $r_t$ and $\Delta\phi = 2\alpha_1/N_t$.
This circuit scales as $\oO(n_t)$.

To encode the velocity grid $v_i$, where $v_{\rm max}\ll 1$ due to the normalization~\eqref{eq:first-normalization-A}, one can use the circuit~\ref{circ:sin}, since $\sin(\phi_i) = \phi_i + \oO(\phi_i^3)$.
In this case, $n_t = n_v$, $\alpha_0 = - v_{\rm max}$, and $\alpha_1 = |\alpha_0| N_v/ (N_v - 1)$.
All nonzero elements of the submatrix $\yC^f$ are encoded by using a single call to the circuit~\ref{circ:sin}.

% ************************************************************************
% *** Submatrix F: profiles ***
% ************************************************************************
\subsection{Encoding the submatrix $\yFP$}\label{sec:FP}
The elements of $\yFP$ described after Eq.~\eqref{eq:FB} depend on $v$ linearly and are encoded using the same technique described in Sec.~\ref{sec:Cf}.
% At the left and right spatial boundaries, these elements have complex values as one can see from Eqs.~\eqref{eq:FL0} and~\eqref{eq:FR0}.
% The total number of these elements is $N_v$.
% The first option is to encoded these elements by using $N_v$ gates $R_c$ according to Eq.~\eqref{eq:Rc}.
% Another option is to encode the real part of these elements by using Eq.~\eqref{eq:sin-gate}, and the imaginary part by using the gate $R_x$. 
% The sum of real and imaginary values can be found by using linear combination of unitaries which requires however an extra ancilla.
% To avoid the extra ancilla and reduce the computation time of the circuit, we use the first option.
% Similarly, we use $\sim N_v$ gates $R_y$ to encode in blocks $\yF^{L,1}$ and $\yF^{R,1}$.
% All blocks $\yF^{B,1}$, which enter the submatrix $\yFP$, can be encoded by a few calls to the circuit similar to~\ref{circ:sin}. 
% The number of this calls does not depend on $n_x$.
% The resulting scaling of the circuit encoding $\yFP$ is $\oO(n_v) + \oO(1)$ where the term $\oO(1)$ appears because after the call to the circuit~\ref{circ:sin}, one needs also to apply $\oO(1)$ STMC gates to compute correct values at the $x$-boundary.

% ************************************************************************
% *** Submatrix CE ***
% ************************************************************************
\subsection{Encoding the submatrix $\yC^E$}\label{sec:CE}
To encode the matrix elements of $\yC^E$ [Eq.~\eqref{eq:CE}], one can use the QSVT to approximate the odd function $v H(v)$.
First of all, the odd function $v H(v)$ is approximated by the following form:
\begin{equation}\label{eq:CE-arcsin}
    f_{vH}(y_i) = p_1 \arcsin(y_i) e^{p_0\arcsin(y_i)^2},
\end{equation}
where $y_i = \sin(\phi_i)$ for $i = [0, N_v)$ where the angles $\phi_i$ are computed using Eq.~\eqref{eq:sin-gate} with the parameters $\alpha_0 = -1$ and $\alpha_1 = |\alpha_0| N_v/(N_v-1)$.
% The parameter $p_0 = -8.0$ does not depend neither on $n_x$ nor $n_v$.
% The parameter $p_1$ decreases linearly with $N_v$ and does not depend on $N_x$ (e.g., $p_1 = -6.045\cdot 10^{-3}$).
The parameter $p_0$ depends neither on $n_x$ nor on $n_v$.
The parameter $p_1$ decreases linearly with $N_v$ and does not depend on $N_x$.
The approximation $f_{vH}(y_i) = v_i H(v_i) \pm \varepsilon_{\rm ce, appr}$ is found by solving a nonlinear least-square problem. 

The function $f_{vH}(y_i)$ is computed by using QSVT, where the variable $y_i$ is encoded by the circuit~\ref{circ:sin}.
Because flat background temperature and density profiles are considered, this function does not depend on $x$ and can be encoded for all $x$-points by a single call to the QSVT circuit.
The absolute error $\epsilon_{\rm CE, qsvt}$ of the QSVT computation of the function~\eqref{eq:CE-arcsin} rapidly decreases with the number of QSVT angles $N_{\rm CE,qsvt}$, e.g. $\epsilon_{\rm CE, qsvt} = 10^{-6}$ for $N_{\rm CE,qsvt} = 16$. 
Each QSVT angle is associated with a single call to the oracle~\ref{circ:sin} that scales linearly with $n_v$.
Thus, the scaling of the circuit encoding the submatrix $\yC^E$ is $\oO(n_v \log_2(\epsilon_{\rm CE, qsvt}^{-1}))$.

If the temperature and density depend on the spatial coordinate, then one needs to implement additional QSVT circuits that would encode the change in the amplitude and the width of the bumps of the function $v H$.
Another option is to substitute these QSVT circuits by a single one by applying Multivariable Quantum Signal Processing (M-QSP)\ycite{Rossi22} which operates with several block-encoding oracles at once, thus, computing a polynomial of multiple variables, i.e. $x$ and $v$ in our case.

% ************************************************************************
% *** Scaling ***
% ************************************************************************
\subsection{Scaling of the BE oracle}\label{sec:scaling}

% % -------------------------------------------------------------------------------------------------------
% % -------------------------------------------------------------------------------------------------------
% \begin{figure}[!t]
% \centering
% \includegraphics[width=0.48\textwidth]{figs/scan-N-gates-OH-fixed.pdf}
% \caption{\label{fig:scan-N-gates-OH-fixed} 
%     The number of STMC gates in the oracle $O_H$ necessary for encoding $\yF_F$ versus the number of nonzero elements in $\yF_F$.
%     For each curve corresponding to a single $n_x$, different markers correspond to various $n_v$, specifically, $n_v = 4,5,6,7$, and $8$, starting from the leftmost marker.
% }\end{figure}
% % -------------------------------------------------------------------------------------------------------
% -------------------------------------------------------------------------------------------------------
\begin{figure}[!t]
\centering
\includegraphics[width=0.48\textwidth]{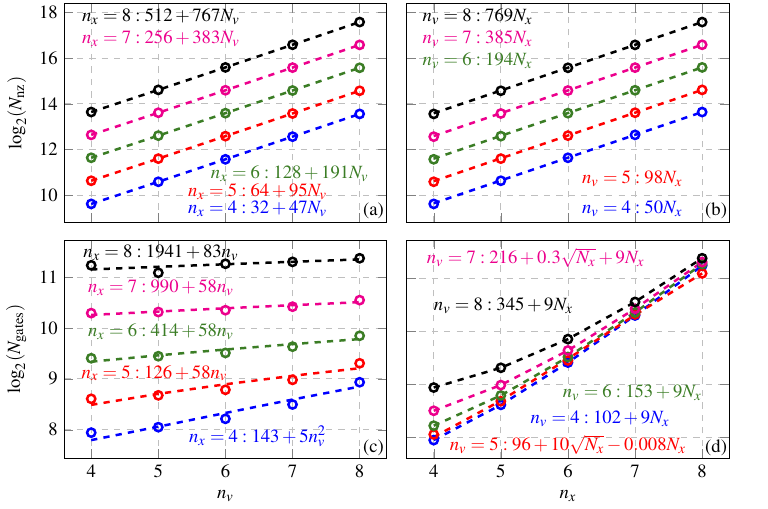}
\caption{
    \label{fig:scan-Ng-Nnz-nx-nv} 
    (a): The dependence of the number of nonzero elements $N_{\rm nz}$ in the matrix $\yF_F$ on $n_v$ for various $n_x$. 
    (b): $N_{\rm nz}$ versus $n_x$ for various $n_v$. 
    (c): The dependence of the number of STMC gates in the oracle $O_H$ necessary for encoding $\yF_F$ on $n_v$ for various $n_x$.
    (d): The dependence of the number of STMC gates on $n_x$ for various $n_v$.
    Here, the text in different colors indicate the fitting equations approximating the scaling with respect to $n_v$ and $n_x$, or $N_v$ and $N_x$.
}
\end{figure}
% -------------------------------------------------------------------------------------------------------
% -------------------------------------------------------------------------------------------------------

The final circuit of the BE oracle encoding the matrix $\yA$ can be found in Ref.~\yocite{EVMcircuit}.
Now, let us estimate the scaling of this circuit.
It comprises the scaling of the oracle $U_D$ and the oracle $O_H$.
As discussed in Sec.~\ref{sec:UD}, $U_D$ scales as $\oO(n_x + n_v)$ due to the arithmetic operators.
The circuit of $O_H$ consists of several pieces: the oracle encoding the submatrix $\yFF$, whose encoding is performed by the procedure discussed in Sec~\ref{sec:FF-oracle}, and the oracles encoding the submatrices  $\yS$ (Sec.~\ref{sec:S}), $\yC^f$ (Sec.~\ref{sec:Cf}), $\yFP$ (Sec.~\ref{sec:FP}), and $\yC^E$ (Sec.~\ref{sec:CE}).
The scaling of the circuit for $\yFF$ is shown in Fig.~\ref{fig:scan-Ng-Nnz-nx-nv} and is estimated as $\oO(\varsigma_{\yFF}{\rm poly}(N_x) {\rm poly}(n_v))$.
The poor scaling with respect to $N_x$ is caused by the matrix elements that appear due to the nonzero diffusivity $\eta$ [Eq.~\eqref{eq:vlasov-diff}].
The nonsparsity $\varsigma_{\yFF}$ of the submatrix $\yFF$ does not change with $N_x$ or $N_v$.
Yet, $\varsigma_{\yFF}$ depends on the discretization method or boundary conditions of the simulated kinetic model. 

The depth of the circuit for $\yS$ scales as $\oO(1)$ if one uses STMC gates, and, according to Ref.~\yocite{Claudon23}, an arbitrary STMC gate controlled by $n$ qubits can be decomposed into $\oO(\log_2(n)^{\log_2(12)}\log_2(1/\epsilon_{\rm STMC}))$ elementary gates, as mentioned in Sec.~\ref{sec:BE-general-idea}.
The oracles for $\yC^f$ and $\yFP$ scale as $\oO(n_v)$.
The scaling of the oracle $\yC^E$ implemented using QSVT is $\oO(n_v\log_2(\epsilon_{\rm CE, qsvt}^{-1}))$.

Then, in summary, the main contribution to the scaling of the BE oracle of the whole matrix $\yA$ is due to the oracle encoding $\yFF$, and the resulting scaling is $\oO[n_x + n_v + \varsigma_{\yFF}{\rm poly}(N_x) {\rm poly}(n_v) + n_v\log_2(\epsilon_{\rm CE, qsvt}^{-1})]$.

\section{Discussion}\label{sec:discussion}

% -----------------------------------------------------------------------
\begin{figure*}[!t]
\centering
\includegraphics[width=0.96\textwidth]{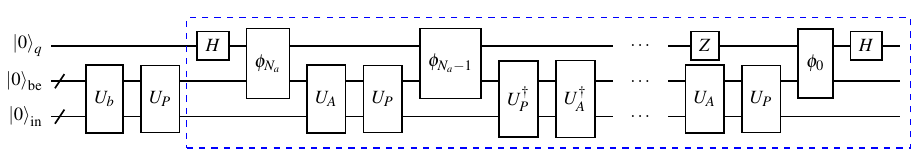}
\caption{
    \label{circ:qsvt-prec} 
    A schematic of the circuit that solves the preconditioned system~\eqref{eq:precond}.
    The oracle $U_P$ encodes the preconditioner $\yP$, the oracle $U_A$ encodes the original matrix $\yA$. 
    The oracle $U_b$ encodes the right-hand-side vector $\yyb$ into the input register `in' which is originally initialized in the zero state.
    The blue box highlights the QSVT circuit which encodes an odd polynomial approximating the inverse matrix $(\yP\yA)^{-1}/\kappa_{PA}$.
    The gates denoted as $\phi_{k}$ correspond to the controlled rotations indicated by the gray dashed boxes in Fig.~\ref{circ:qsvt}.
}
\end{figure*}
% -----------------------------------------------------------------------

\subsection{Preconditioning}
As mentioned earlier, a QSVT-based algorithm for solving Eq.~\eqref{eq:axb} essentially amounts to inverting the matrix $\yA$. 
Unfortunately, for boundary-value wave problems, and kinetic problems in particular, $\yA$ typically has large condition numbers $\kappa_{A}$.
This makes accurate inversion challenging and also complicates extracting classical information, since the success probability of the circuit scales as $O(\kappa_A^{-1})$\ycite{Novikau23}. 
A solution to this can be in using a preconditioner.
Specifically, suppose one finds a matrix $\yP$ such that the matrix $\yP\yA$ has a much smaller condition number than that of $\yA$, $\kappa_{PA} \ll \kappa_{A}$. 
From Eq.~\eqref{eq:axb}, it follows that
\begin{equation}\label{eq:precond}
    \yP\yA \psi = \yP \yyb.
\end{equation}
Then, the matrix $\yP\yA$ is easier to invert than the original matrix $\yA$, so it may be possible to easily calculate the solution in the form
\begin{equation}\label{eq:psi-prec}
    \psi = (\yP\yA)^{-1}\yP\yyb.
\end{equation}
A schematic of the QSVT circuit with a preconditioner is shown in Fig.~\ref{circ:qsvt-prec}.
Here, $\yP$ is computed by the separate BE oracle $U_P$, the oracles $U_A$ and $U_b$ can be the same as in the problem without a preconditioner, so the only additional step needed is to implement $U_P$. 
An advantage of this approach (compared to constructing oracles for $\yP\yA$ and $\yP\yyb$) is that $U_P$ does not have to be constructed precisely. 
If, instead of the intended $\yP$, $U_P$ encodes a slightly different preconditioner $\yP'$, this changes Eq.~\eqref{eq:precond} into $\yP'\yA \psi = \yP'\yyb$, but the latter is still equivalent to the original Eq.~\eqref{eq:axb} leading to an equivalent solution $\psi = (\yP'\yA)^{-1}\yP'\yyb$. 
% that simply modifies Eq.~\eqref{eq:precond} but not Eq.~\eqref{eq:axb}, leading to an equivalent solution $\psi = (\yP'\yA)^{-1}\yP'\yyb$. 
As long as the condition number of $\yP'\yA$ is comparable to that of $\yP\yA$, the matrix $\yP'$ serves the role of a preconditioner just as well as the matrix $\yP$ and thus the modified $U_{P}$ is just as good as the intended $U_P$.

The implementation of the combined operator $U_P U_A$ may require a simple compression gadget\ycite{Fang23} to guarantee correct computation of the product $\yP'\yA$. 
This gadget will need two ancillae, two quantum decrementors and the adder $A2$ (Fig.~\ref{fig:adder2}), i.e. the arithmetic operators that scale linearly with the number of qubits in the gadget.
Thus, the gadget will not deteriorate the overall algorithm scaling.

\subsection{Remaining challenges}

% To extract classical information, such as the electric field energy or field spectrum, from this quantum state, one can use the techniques discussed, for example, in Ref.~\yocite{Novikau23}.
% The measurement will require at least $\oO(\kappa_{PA})$ repetitions of the circuit~\ref{circ:qsvt-prec}.

The main issues that remain to be addressed in future works are the following.
% The first one is the development of an algorithm similar to that presented in Sec.~\ref{sec:OH} for constructing the oracle $U_D$.
% In this work, this oracle was constructed manually ad hoc as explained in Sec.~\ref{sec:UD}.
First, the manual construction of an oracle for $U_D$ that we proposed in Sec.~\ref{sec:UD} will need to be replaced with an automated procedure.
The algorithm may be based on the same foundations discussed in Sec.~\ref{sec:OH}.
% However, to build $U_D$, one is interested only in the row and column indices, $i_r$ and $i_c$, of nonzero matrix elements in $\yA$ instead of the values of the matrix elements.
% Therefore, $\yA$ can be decomposed again into sets where each set now stores column indices $i_c$ (or the shifts $i_c - i_r$) of all nonzero elements in the rows $i_r$.
% The final goal of this algorithm should be the encoding of $i_c$ or $i_c - i_r$ into ancilla qubits by a polylogarithmic number of STMC gates.

Another issue, which has already been mentioned in Sec.~\ref{sec:initialization}, is that the encoding of the source with a nontrivial profile may significantly reduce the success probability of the overall quantum algorithm.
For instance, both QSVT or QETU methods compute a Gaussian with success probability scaling as $\oO(\sigma_G/x_{\rm max})$, where $\sigma_G$ is the width of the Gaussian\ycite{Kane23}.
To increase the probability, it may be better to approximate the source profile with a different function that is easier to encode. 
% It might be that the easiest way to increase the probability is to approximate the source by a rough profile that could be encoded more efficiently.
For instance, the encoding of a strongly localised source (i.e. $\delta$-function) requires one or several Pauli $X$ gates and has the success probability equal to unity.
An impulse with the amplitude $2^{-n_s/2}$ potentially can be encoded using $n_s$ Hadamard gates and $\oO(n_s)$ Pauli $X$ gates.
Hence, it may be possible to approximate a necessary source by a set of pulses, which then can be encoded efficiently and ensure high success probability at the same time.

% efficiently encoded with a high success probability.

\section{Conclusions}\label{sec:conclusions}
% In this paper, we developed a numerical and analytical description of stationary dissipative waves in electrostatic kinetic electron uniform plasma driven by an oscillating source.
% This stationary kinetic problem was formulated in the form of a system of linear equations.
% Results from its numerical modeling by a classical parallelized sparse method agree with the corresponding analytical description. 
% After that, we developed several techniques necessary for the block encoding of the kinetic model into a quantum circuit.
% The matrix describing the kinetic wave system was split into several submatrices where each submatrix is encoded in its own way.
% The submatrix which includes spatial and velocity derivatives have the most complex structure.
% It was proposed to represent this submatrix as a group of sets be combining equal matrix elements.
% In this way, the submatrix is turned into a compressed form which is easier to block encode in an efficient manner thus significantly improving the scaling of the resulting BE oracle.
% Yet, the main reason still deteriorating the BE scaling is intersecting sets where matrix elements in the overlapped regions should be corrected by additional STMC gates.
% The proposed algorithm can be further improved by including machine learning techniques and can serve as a foundation for the development of more advanced methods for the block encoding complex matrices used in classical plasma and fluid multidimensional problems.

In this paper, we propose an algorithm for encoding of linear kinetic plasma problems in quantum circuits. 
The focus is made on modeling of electrostatic linear waves in one-dimensional Maxwellian electron plasma. 
The waves are described by the linearized Vlasov–Amp\`ere system with a spatially localized external current that drives plasma oscillations. 
This system is formulated as a boundary-value problem and cast in the form of a linear vector equation~\eqref{eq:axb}, to be solved using the QSP-based algorithm. 
The latter requires encoding of the matrix $\yA$ in a quantum circuit as a subblock of a unitary matrix. 
We propose an algorithm for encoding $\yA$ into a circuit using a compressed form of the matrix. 
This significantly improves the scaling of the resulting BE oracle with respect to the velocity coordinate. 
However, further thorough analysis is required to improve the scaling along the spatial coordinate.
The proposed algorithm can serve as a foundation for developing BE algorithms in more complex kinetic linear plasma problems, for example, for modeling of electromagnetic waves in magnetized plasma.

\begin{acknowledgments}
The authors thank Ilon Joseph for valuable discussions.
The research described in this paper was supported by the Laboratory Directed Research and Development (LDRD) Program at Princeton Plasma Physics Laboratory (PPPL), a national laboratory operated by Princeton University, and by the U.S. Department of Energy (DOE) Office of Fusion Energy Sciences “Quantum Leap for Fusion Energy Sciences” Project No. FWP-SCW1680 at Lawrence Livermore National Laboratory (LLNL). 
Work was performed under the auspices of the U.S. DOE under PPPL Contract DE-AC02-09CH11466 and LLNL Contract DE-AC52–07NA27344.
\end{acknowledgments}

%% -------------------------------------------------------
%% --- Appendix ---
%% -------------------------------------------------------
\appendix
\section{Analytical solution}\label{app:analytical}

Here, we derive an analytical solution of Eq.~\eqref{sys:VM-interm} for the special case when the plasma is homogeneous ($n = T = 1$).
By applying the Laplace transform (Appendix~\ref{app:transform}) to Eq.~\eqref{sys:VM-interm}, we obtain:
\begin{subequations}\label{eqs:after-Laplace}
\begin{eqnarray}
    &&\yi\omega\yLf - v\partial_x\yLf + \yLE\partial_v \yBF = 0,\\
    &&\yi\omega\yLE + \int v \yLf \diff v = -\yLS
\end{eqnarray}
\end{subequations}
[assuming $\ypf_0 \equiv \ypf(t=0) = 0$], where $\yLS = \ypE_0 - \yLj$ due to Eq.~\eqref{eq:Laplace-dt}.
Let us also apply the Fourier transform in space (Appendix~\ref{app:transform}), assuming zero boundary conditions at infinity (i.e. $\yLf|_{x\to\pm\infty} = \yLE|_{x\to\pm\infty} = 0$).
This leads to
\begin{subequations}\label{eqs:after-Laplace-Fourier}
\begin{eqnarray}
 &&\yi\omega\yLFf - \yi k v \yLFf + \yLFE\partial_v \yBF = 0,\\
 &&\yi\omega\yLFE + \int v \yLFf\diff v = -\yLFS.
\end{eqnarray}
\end{subequations}
From the above equations, one gets that
\begin{subequations}
\begin{eqnarray}
    &&\yLFf = - \frac{\yi\yLFE}{k} \frac{\partial_v \yBF}{v - \omega/k},\\
    &&\yLFE = \yi\yLFS\left(\omega - \frac{1}{k} \int \frac{v\partial_v \yBF}{v - \omega/k}\diff v \right)^{-1}.\label{eq:E-w-k-general}
\end{eqnarray}
\end{subequations}
For the background Maxwellian distribution~\eqref{eq:maxwellian}, the electric field becomes
\begin{subequations}
\begin{eqnarray}
    &&\yLFE              = \frac{\yi\yLFS}{\omega\epsilon(\omega,k)},\label{eq:E-LF-1}\\
    &&\epsilon(\omega,k) = 1 + [1 + \xi Z_0(\xi)]/k^2,
\end{eqnarray}
\end{subequations}
where $\xi = \omega/(k\sqrt{2})$, and the function $Z_0(\xi)$ is defined via the plasma dispersion function $Z(\xi) = \sqrt{\pi}e^{-\xi^2}[\yi - {\rm erfi}(\xi)]$ as\ycite{Stix92}
\begin{equation}
Z_0(\xi) = 
    \left\{ 
    \begin{aligned}
        Z(\xi),\quad k > 0,\\
        -Z(-\xi),\quad k < 0.
    \end{aligned} 
    \right.
\end{equation}

By applying the Laplace transform in time and the Fourier transform in space to Eq.~\eqref{eq:charge-conserv}, one also obtains:
\begin{equation}
    -\yi\omega\yLFrho + \yi k \yLFj = \yFrho,
\end{equation}
where $\yFrho = \yrhos_k(t = 0)$,
whilst the initial electric field satisfies the Gauss' law:
\begin{equation}
    \yi k \yFEi = \yFrho.
\end{equation}
Thus, one obtains
\begin{equation}
    \yFEi = \yLFj - \omega\yLFrho/k,
\end{equation}
from where one can see that 
\begin{equation}
    \yLFS = -\omega\yLFrho/k.
\end{equation}
As a result, Eq.~\eqref{eq:E-LF-1} yields:
\begin{equation}\label{eq:E-LF-2}
    \yLFE = - \frac{\yi\yLFrho}{k\epsilon(\omega,k)}.
\end{equation}
% This means that we define the source by specifying the source charge density.
Let us assume an oscillating source charge density:
\begin{equation}
    \yrhos(t,x) = Q(x)e^{-\yi\omega_0 t}.
\end{equation}
This corresponds to
\begin{equation}
    \yLFrho = \frac{i\yFQ}{\omega-\omega_0},
\end{equation}
and Eq.~\eqref{eq:E-LF-2} yields
\begin{equation}
    \yLFE = \frac{\yFQ}{k(\omega-\omega_0)\epsilon(\omega,k)}.
\end{equation}
To find the evolution of the electric field in time, we perform the inverse Laplace transform [Eq.~\eqref{eq:inv-L-residue}]:
\begin{equation}
    \yFE = -\frac{\yi\yFQ}{k} \left[
        \frac{e^{-\yi\omega_0 t}}{\epsilon(\omega_0,k)} + 
        \sum_{q\geq 1}\frac{e^{-\yi\omega_q t}}{(\omega_q - \omega_0)\partial_{\omega}\epsilon(\omega_q,k)} 
    \right].
\end{equation}
We are interested only in established spatial distribution of the electric field that is observed at $t\to +\infty$.
Because of the Maxwellian background distribution~\eqref{eq:maxwellian}, the plasma is stable, i.e. ${\rm Im}\,\omega_q(k)\leq 0$, for all $q \geq 1$. 
Therefore, $e^{-\yi\omega_q t}\to 0$ at $t\to +\infty$, for all such $q$, and the Fourier components of the electric field become
\begin{equation}\label{eq:Ek}
    \yFE = -\frac{\yi\yFQ}{k} \frac{e^{-\yi\omega_0 t}}{\epsilon(\omega_0,k)}.
\end{equation}
We use the source spatial distribution as in Eq.~\eqref{eq:charge-source-x}, whose Fourier transform is
\begin{equation}
    \yFQ = \yi k \sqrt{2\pi}\Delta_S e^{-k(\Delta_S^2 k + 2\yi x_0)/2}.
\end{equation}
Finally, to find $\ypE(x)$ at $t\to +\infty$, we need to compute the inverse Fourier transform of $\yFE$:
\begin{equation}\label{eq:E-analytical}
    \ypE(x) = -\frac{\yi}{2\pi} \int_{-\infty}^{+\infty}\frac{\yFQ e^{\yi k x}}{k\epsilon(\omega_0,k)}\diff k,
\end{equation}
that can be done numerically. 
In Sec.~\ref{sec:classical-results}, this result is used for benchmarking our classical numerical simulations.

% ------------------------------------------------------------------------------------
% ------------------------------------------------------------------------------------
\section{Laplace and Fourier transformation}\label{app:transform}
For a given function $y(t)$, we define the temporal Laplace transform as
\begin{equation}\label{eq:Laplace-t}
    y_\omega = \mathcal{L}[y(t)] \equiv \int_0^{+\infty} y(t) e^{\yi \omega t} \diff t,
\end{equation}
where $\omega$ is a complex value such that ${\rm Im}\,\omega$ is large enough for the integral to converge. 
Accordingly, the Laplace transform of the time derivative of $y$ can be written as
\begin{equation}\label{eq:Laplace-dt}
    \int_0^{+\infty} e^{\yi \omega t} \partial_t y  \diff t = -\yi\omega y_\omega - y_0,
\end{equation}
where we integrated by parts and introduced $y_0 = y(t=0)$.
The Laplace transforms of higher-order derivatives are derived similarly.

The inverse Laplace transform is 
\begin{equation}
    y(t) = \mathcal{L}^{-1}[y_\omega] \equiv \frac{1}{2\pi} \int_{-\infty + \yi\gamma}^{+\infty+\yi\gamma} y_\omega e^{-\yi\omega t}\diff \omega.
\end{equation}
In the inverse transformation, the integration contour goes along the axis $(-\infty +\yi \gamma,+\infty + \yi\gamma)$ with a positive $\gamma$ such that all singularities of $y_\omega$ lie below the axis.
Provided that $y_\omega$ is well-behaved at ${\rm Im}\,\omega \to -\infty$ and also that the only singularities of $y_\omega$ are poles $\omega_j$, one can shift the integration contour downward in the complex-$\omega$ plane while still encircling the poles from above. 
Then, the contribution from the horizontal part of the contour vanishes and only the pole contributions remain.
In this case, the integral can be computed as a sum of residues at the function poles:
\begin{equation}\label{eq:inv-L-residue}
    \mathcal{L}^{-1}[y_\omega] = -\yi \sum_{j} {\rm Res}\ylb y_\omega e^{-\yi\omega t}, \omega_j\yrb.
\end{equation}

We define the Fourier transform of a given function $z(x)$ as
\begin{equation}
    z_k = \int_{-\infty}^{+\infty} z(x) e^{-\yi k x}\diff x,
\end{equation}
where $k$ is a real value.
The inverse Fourier transform is then
\begin{equation}
    z(x) = \frac{1}{2\pi} \int_{-\infty}^{+\infty} z_k e^{\yi k x}\diff k.
\end{equation}

\section{Supplemental gates}\label{app:sup-gates}
To encode the complex value $e_c = |e_c| \exp[\yi\arg(e_c)]$, we use the following operator:
\begin{equation}\label{eq:Rc}
    R_c(\theta_z, \theta_y) = R_y(\theta_y) R_z(\theta_z),
\end{equation}
with $\theta_z = - 2 \arg(e_c)$ and $\theta_y = 2 \arccos(|e_c|)$.

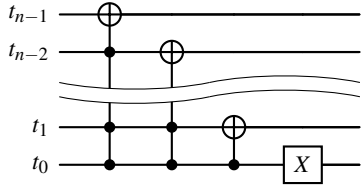
\begin{figure}[!t]
\centering
\begin{quantikz}[row sep={0.5cm,between origins}]
\lstick{$t_{n-1}$} &\targ{}   &\qw       &\qw       &\qw      &\qw  \\
\lstick{$t_{n-2}$} &\ctrl{-1} &\targ{}   &\qw       &\qw      &\qw  \\
\wave          &          &          &          &         &     \\
\lstick{$t_1$} &\ctrl{-2} &\ctrl{-2} &\targ{}   &\qw      &\qw  \\
\lstick{$t_0$} &\ctrl{-1} &\ctrl{-1} &\ctrl{-1} &\gate{X} &\qw
\end{quantikz}
\caption{\label{fig:incrementor} The circuit of an incrementor, denoted as $A1$, which acts on the target register $t$ with $n$ qubits. 
The circuit of a decrementor denoted as $S1$ is inverse to the circuit shown here.}
\end{figure}

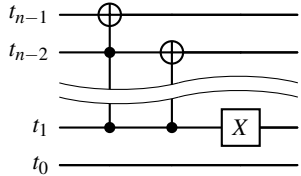
\begin{figure}[!t]
\centering
\begin{quantikz}[row sep={0.5cm,between origins}]
\lstick{$t_{n-1}$} &\targ{}   &\qw       &\qw       &\qw \\
\lstick{$t_{n-2}$} &\ctrl{-1} &\targ{}   &\qw       &\qw \\
\wave          &          &              &          &    \\
\lstick{$t_1$} &\ctrl{-2} &\ctrl{-2}     &\gate{X}  &\qw \\
\lstick{$t_0$} &\qw       &\qw           &\qw       &\qw
\end{quantikz}
\caption{\label{fig:adder2} The circuit of an adder, denoted as $A2$, which adds $2$ to the unsigned integer encoded in the target register $t$ with $n$ qubits. 
The circuit of a subtractor by $2$, $S2$, is inverse to the circuit shown here.}
\end{figure}

\begin{figure}[!t]
\centering
\begin{quantikz}[row sep={0.5cm,between origins}]
\lstick{$t_{n-1}$} &\targ{}   &\qw      &\qw      &\qw      &\qw      &\qw      &\qw      &\targ{}   &\qw \\
\lstick{$t_{n-2}$} &\ctrl{-1} &\targ{}  &\qw      &\qw      &\qw      &\qw      &\targ{}  &\ctrl{-1} &\qw \\
\wave              &          &         &         &         &         &         &         &          &    \\
\lstick{$t_3$}     &\ctrl{-2} &\ctrl{-2}&\targ{}  &\qw      &\qw      &\targ{}  &\ctrl{-2}&\ctrl{-2} &\qw \\
\lstick{$t_2$}     &\ctrl{-1} &\ctrl{-1}&\ctrl{-1}&\gate{X} &\targ{}  &\ctrl{-1}&\ctrl{-1}&\ctrl{-1} &\qw \\
\lstick{$t_1$}     &\qw       &\qw      &\qw      &\targ{}  &\ctrl{-1}&\ctrl{-1}&\ctrl{-1}&\ctrl{-1} &\qw \\
\lstick{$t_0$}     &\qw       &\qw      &\gate{X} &\ctrl{-1}&\ctrl{-1}&\ctrl{-1}&\ctrl{-1}&\ctrl{-1} &\qw
\end{quantikz}
\caption{\label{fig:adder3} The circuit of an adder, denoted as $A3$, which adds $3$ to the unsigned integer encoded in the target register $t$ with $n$ qubits. 
The circuit of a subtractor by $3$, $S3$, is inverse to the circuit shown here.}
\end{figure}
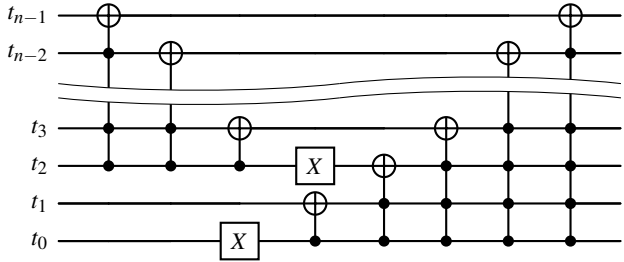

To shift an unsigned integer encoded in qubits by an integer more than $1$, one can use the gates discussed in Refs.~\yocite{Novikau23, Suau21, Draper00}.
However, since here we need only shifts by $\pm 1$, $\pm 2$, and $\pm 3$, we introduce shorter circuits for these operations.
The shift by $\pm 1$ corresponds to an incrementor or decrementor (Fig.~\ref{fig:incrementor}). 
These operators are denoted as $A1$ and $S1$, correspondingly, in Fig.~\ref{circ:OM}.
The shift by $\pm 2$ does not modify the least significant qubit and, therefore, corresponds to the incrementor or decrementor whose circuit is shifted upwards (Fig.~\ref{fig:adder2}).
These operators are denoted as $A2$ and $S2$, correspondingly.
The circuit for the adder by $3$, shown in Fig.~\ref{fig:adder3}, can be understood as a combination of the circuits of an adder by $4$ and a subtractor by $1$.
The adder and subtractor by $3$ are denoted as $A3$ and $S3$, correspondingly.
The number of STMC gates in all these operators scales as $\oO(n)$.

% ---------------------------------------------------------------------------------------------
% ---------------------------------------------------------------------------------------------
% ---------------------------------------------------------------------------------------------
% ---------------------------------------------------------------------------------------------
% ---------------------------------------------------------------------------------------------
% ---------------------------------------------------------------------------------------------

\bibliography{main}

\begin{thebibliography}{10}
\bibitem{Notea}
In this paper, matrices and vectors are indicated by bold symbols, as in Eq.~\protect \eqref {eq:axb}, while quantum-state vectors, oracles, and operators in quantum circuits are denoted with non-bold italic symbols, e.g. $U$ and $\psi $.
\bibitem{Noteb}
Although the term `sparsity' is commonly used in literature, it is better to use the term ``nonsparsity'' when one refers to the parameter $\varsigma $ which grows than a matrix becomes less sparse. In other words, matrices with a large $\varsigma $ are characterized by low sparsity.
\end{thebibliography}

\end{document}